\documentclass[aps,prl,twocolumn,superscriptaddress,amsmath,amssymb,citeautoscript]{revtex4}

\usepackage{graphicx}
\usepackage{dcolumn}
\usepackage{booktabs}
\usepackage{bm}
\usepackage{color}
\usepackage{multirow}

\newcommand{\Tc}{\ensuremath{T_{\rm c}}}
\newcommand{\gn}{\ensuremath{\gamma_{\rm n}}}
\newcommand{\gs}{\ensuremath{\gamma_{\rm s}}}

\newcommand{\cel}{\ensuremath{c_{\rm el}}}

\newcommand{\EF}{\ensuremath{E_{\rm F}}}
\newcommand{\SITS}{Sn$_{1-x}$In$_{x}$Te$_{1-y}$Se$_{y}$}
\newcommand{\SIT}{Sn$_{1-x}$In$_{x}$Te}

\newcommand{\PTT}{Pb$_{1-x}$Tl$_{x}$Te}

\begin{document}

\title{Tailoring band-structure and band-filling in a simple cubic \\(IV, III)\,--\,VI superconductor}

\author{M.~Kriener$^{\dagger}$}
\email[corresponding author: ]{markus.kriener@riken.jp}
\author{M.~Kamitani}
\thanks{These three authors contributed equally.}
\affiliation{RIKEN Center for Emergent Matter Science (CEMS), Wako 351-0198, Japan}
\author{T.~Koretsune}
\thanks{These three authors contributed equally.}
\affiliation{RIKEN Center for Emergent Matter Science (CEMS), Wako 351-0198, Japan}
\affiliation{Department of Physics, Tohoku University, Miyagi 980-8578, Japan}
\author{R.~Arita}
\author{Y.~Taguchi}
\affiliation{RIKEN Center for Emergent Matter Science (CEMS), Wako 351-0198, Japan}
\author{Y.~Tokura}
\affiliation{RIKEN Center for Emergent Matter Science (CEMS), Wako 351-0198, Japan}
\affiliation{Department of Applied Physics and Quantum-Phase Electronics Center (QPEC), University of Tokyo, Tokyo 113-8656, Japan}

\date{\today}

\begin{abstract}
Superconductivity and its underlying mechanisms are one of the most active research fields in condensed-matter physics. An important question is how to enhance the transition temperature \Tc\ of a superconductor. In this respect, the possibly positive role of valence-skipping elements in the pairing mechanism has been attracting considerable interest. Here we follow this pathway and successfully enhance \Tc\ up to almost 6~K in the simple chalcogenide SnTe known as topological crystalline insulator by doping the valence-skipping element In and codoping Se. A high-pressure synthesis method enabled us to form single-phase solid solutions \SITS\ over a wide composition range while keeping the cubic structure necessary for the superconductivity. Our experimental results are supported by density-functional theory calculations which suggest that even higher \Tc\ values would be possible if the required doping range were experimentally accessible. 
\end{abstract}

\maketitle

Narrow-gap chalcogenide semiconductors like GeTe, PbTe, or Bi$_2$Se$_3$ have attracted long-lasting interest due to their surprisingly rich variety on physical properties given their chemical simplicity. Also, the abundance of interesting features can be greatly enhanced by doping. In recent years, this class of materials has become even better known since many among them were found to host topological insulator phases of matter where the bulk is insulating while the surface allows metallic conduction owing to a peculiar band structure and strong spin-orbit interaction \cite{hasan10a,qi11a,ando15a}. One prominent example is SnTe, which was predicted and soon after experimentally verified to be a topological crystalline insulator \cite{hsieh12b,ytanaka12b}, where the topological nontrivial band structure is protected by the mirror symmetry of the underlying crystal structure \cite{fu11a}. SnTe, or more precisely Sn$_{1-\delta}$Te, is also a self-doped superconductor with a superconducting transition temperature $\Tc<300$~mK \cite{hein65a}. However when doping In, \Tc\ is enhanced by one order of magnitude \cite{nakajima73a,bushmarina86a,erickson09a}.

This enhancement and the discovery of the topological nature of SnTe have generated considerable interest in this system in recent years \cite{sasaki12a,tsato13a,novak13a,rdzhong13a,lphe13a,maurya14a,saghir14a,hashimoto15a,haldolaarachchige16a,maeda17b}. A zero-bias conduction peak was found in point-contact spectroscopy experiments on \SIT\ at low doping $x \approx 0.045$ \cite{sasaki12a}. In addition, ARPES measurements confirmed that the topological band structure survives against the doping \cite{tsato13a}, and it was concluded that \SIT\ is a promising candidate to realize topological superconductivity where the superconducting gap function possesses a nontrivial topology. By contrast, a recent nuclear-magnetic-resonance study on similarly low-doped \SIT\ suggests conventional superconductivity \cite{maeda17b}. 

All these works focus on $x \leq 0.5$, which is the solubility limit of In in cubic SnTe at ambient conditions. The end member InTe is a tetragonal semiconductor and does not superconduct. However, when synthesizing InTe under a pressure of $p\sim 3$~GPa, cubic InTe with rock-salt structure forms and is metastable at room temperature. Moreover it superconducts below $\Tc \sim 3$~K \cite{boemmel63a,darnell63a}, motivating this study to synthesize \SIT\ for $x \geq 0.5$ and their Se-codoped analogues by employing a high-pressure synthesis method. 

Polycrystalline samples with $x \geq 0.5$were prepared by a high-pressure technique at 5~GPa and $1200 - 1300^{\circ}$C. For comparison, we also synthesized samples for $x<0.5$ by conventional melt growth and confirmed quantitative agreement with the results found in literature, e.g., that \Tc\ increases roughly linearly for $0.1<x<0.5$ \cite{novak13a,rdzhong13a,lphe13a,maurya14a,saghir14a,hashimoto15a,haldolaarachchige16a,maeda17b}. The synthesis conditions are comparatively summarized in Table~S1 in the Supplemental Material \cite{Suppl}. Resistivity and specific heat were measured in a commercially available system (PPMS, Quantum Design) by a standard four-probe technique and a relaxation method, respectively. The electronic structures, phonon frequencies, and electron-phonon couplings were calculated in the framework of the density-functional theory (DFT) and the density functional perturbation theory as implemented in the quantum-ESPRESSO package \cite{giannozzi09a}. Then, theoretical \Tc\ values were obtained using the McMillan-Allen-Dynes formula \cite{allen75a}. For details, see Section~S6 in the Supplemental Material \cite{Suppl}.


\begin{figure}[t]
\centering
\includegraphics[width=7.5cm,clip]{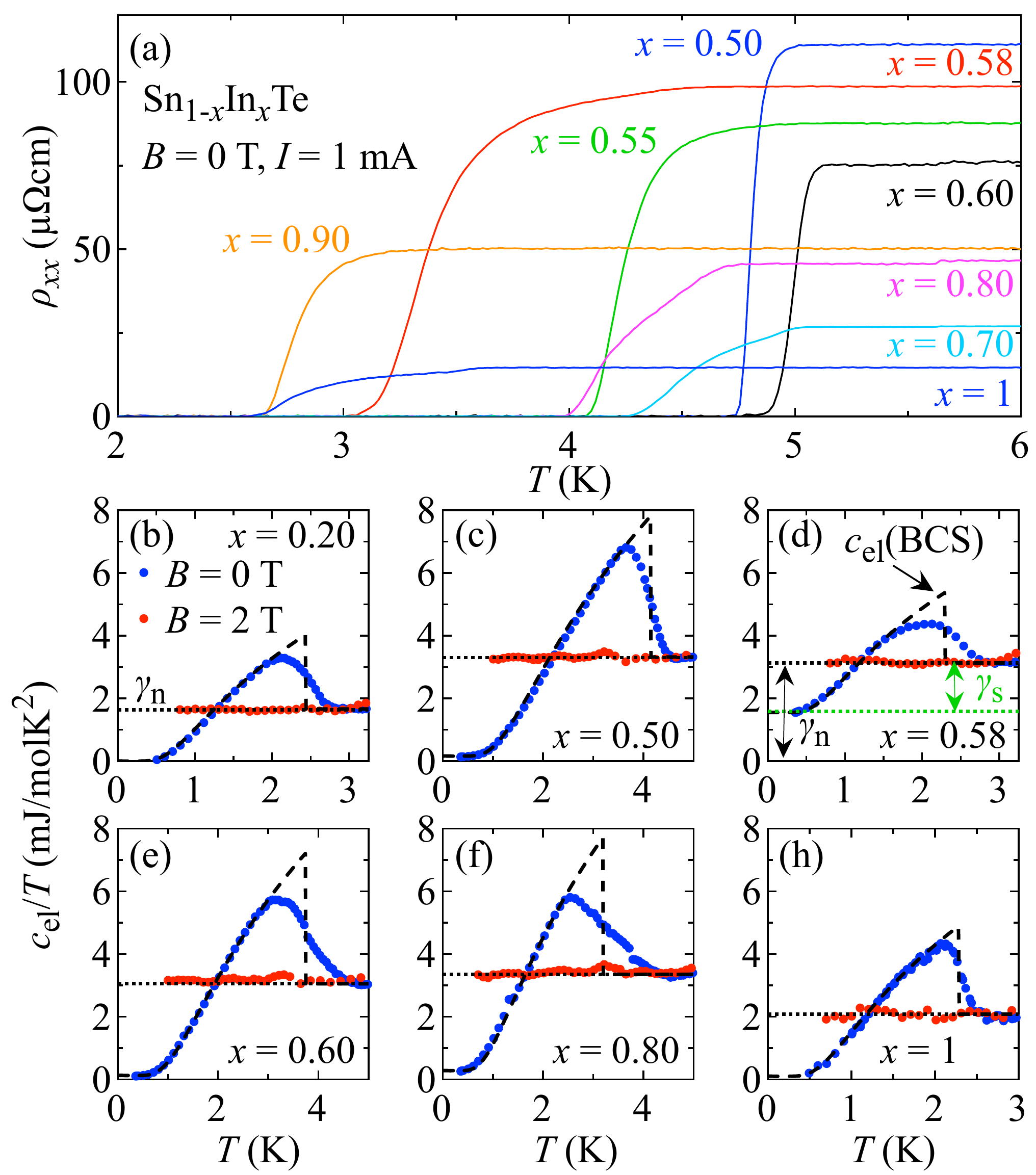}
\caption{(a) Temperature-dependent resistivity data of selected samples $0.5 \leq x \leq 1$. (b) to (h) Electronic specific-heat data $\cel/T$ in $B=0$~T (blue data) and 2~T (red) is plotted against temperature ($T$) for selected samples $0.2 \leq x \leq 1$. A magnetic field of 2~T is sufficient to suppress the superconductivity in this system. The black dotted lines denote the electronic specific-heat coefficient of the normal state \gn. The green dotted line in (d) indicates the residual density of states, and its difference to \gn\ corresponds to the superconducting density of states \gs\ (not shown for the other samples, see text). Dashed lines are modeled BCS electronic specific heat, see Section~S5 in the Supplemental Material \cite{Suppl} for details.}
\label{fig1}
\end{figure}
Temperature-dependent resistivity $\rho_{xx}$ data of selected high-pressure grown samples of \SIT\ ($0.5 \leq x \leq 1$) are summarized in Fig.~\ref{fig1}(a). All examined materials exhibit superconducting transitions between 2~K and 5~K. One remarkable feature is the unexpected and steep suppression of $\Tc = T(\rho_{xx} = 0)$ in the narrow doping range around $x\approx 0.58$, which was confirmed to be quite reproducible for several samples from different synthesis runs. For $x =0.525$ (data not shown), we find $\Tc \approx 4.75$~K which decreases down to the minimum-\Tc\ of 3.1~K for $x=0.58$, amounting to a suppression of $\sim 35$\%. Interestingly, \Tc\ adopts its maximum value 4.9~K for $y=0$ at a merely slightly higher In concentration of $x=0.6$. Upon further doping, \Tc\ monotonously decreases towards InTe. 

Figures~\ref{fig1}(b)\,--\,(h) show superconducting and normal-state electronic specific-heat data \cel\ of selected samples $0.2 \leq x \leq 1$ displayed as $\cel/T$ vs $T$ (blue data: $B=0$~T, red: 2~T). The dotted horizontal lines denote the respective electronic specific-heat (Sommerfeld) coefficients \gn. The dashed lines represent the Bardeen-Cooper-Schrieffer (BCS) electronic specific heat for the experimental values of \Tc\ and \gs\ (see Section~S5 in the Supplemental Material [\onlinecite{Suppl}] for their exact definition and the details of the specific-heat analysis). Concurrently with the suppression of \Tc, the superconducting volume fraction also decreases drastically to roughly 50\% for $x=0.58$ as indicated by the residual density of states (green dotted line) in Fig.~\ref{fig1}(d). For all other samples the specific-heat analysis yielded superconducting volume fractions of 90 -- 100\%, indicating the bulk nature of the superconductivity in this system. We also find that \cel\ can be satisfactorily described in a weak-coupling BCS scenario throughout the doping series as indicated by the dashed lines in each panel. However, at low doping concentrations $x \leq 0.5$, the description is slightly better when assuming a more strong-coupling BCS scenario in agreement with earlier studies \cite{novak13a,haldolaarachchige16a}, see also Fig.~S6 in the Supplemental Material \cite{Suppl}.
Magnetization measurements also confirm large shielding signals (cf.\ Fig.~S3 in the Supplemental Material \cite{Suppl}).
\begin{figure}[t]
\centering
\includegraphics[width=8.5cm,clip]{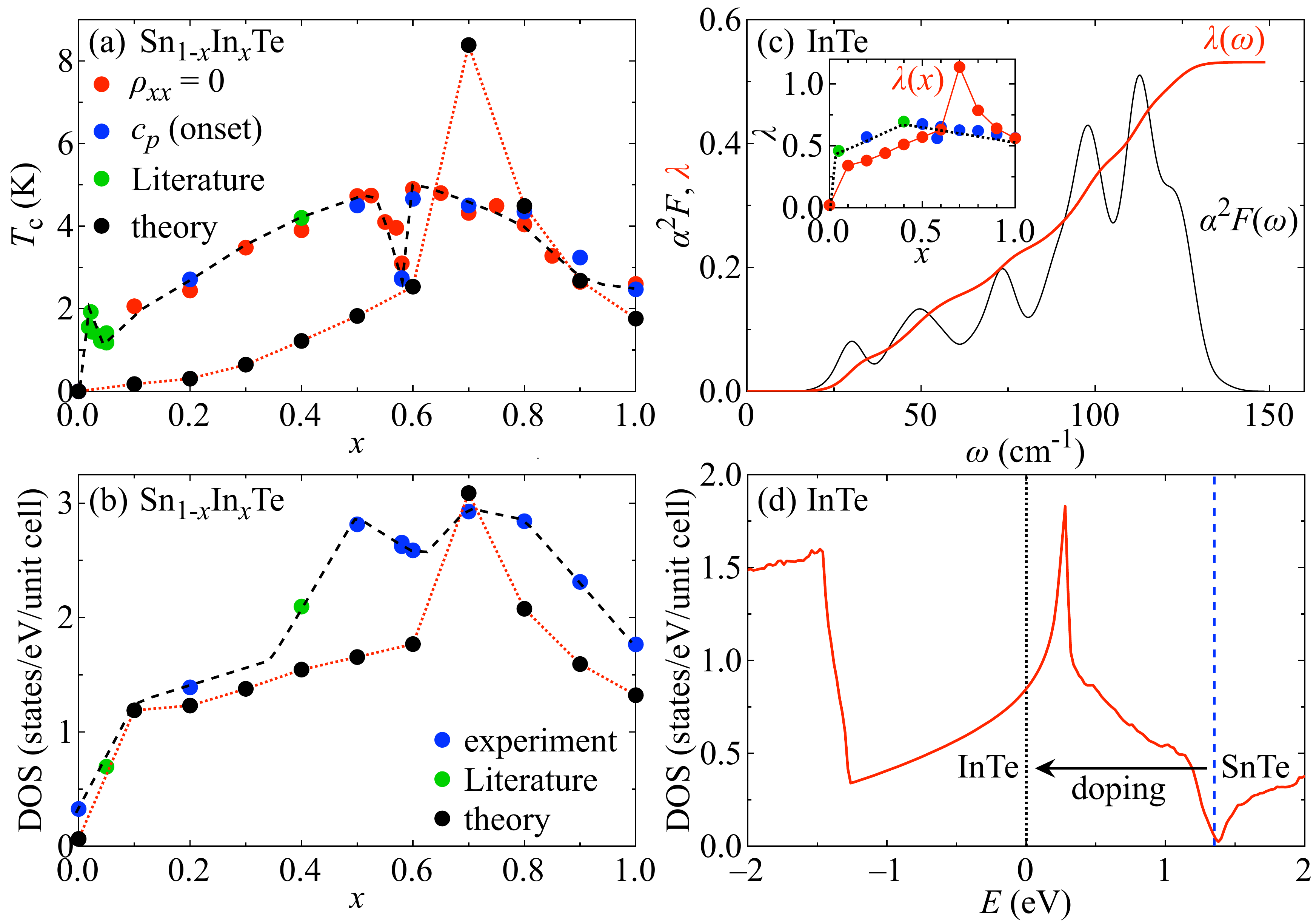}
\caption{(a) Superconducting \Tc\ vs In concentration $x$: Red data points correspond to the temperatures at which zero resistance is observed, blue to the onset temperature of the jump-like anomaly in specific-heat data and black data points are calculated \Tc\ values. (b) Density of states DOS vs $x$: Blue data points were estimated from the experimental electronic specific-heat coefficient \gn\ and black data points are theoretical DOS values (see text for details). Green data points in (a) and (b) are taken from literature (Refs.~\onlinecite{novak13a} and \onlinecite{haldolaarachchige16a}) for comparison. The dashed and dotted lines in both panels are guides to the eyes. (c) Calculated Eliashberg function $\alpha^2 F$ (black) and (integrated) electron-phonon coupling constant $\lambda$ (red) as a function of the phonon frequency $\omega$ for InTe. Inset shows the $x$ dependence of calculated $\lambda$ values (red symbols) compared with those estimated from specific-heat data (blue). Green data points are taken from Ref.~\onlinecite{haldolaarachchige16a}. (d) Calculated DOS for InTe as a function of energy. The Fermi energy \EF\ of InTe is defined as 0 and indicated with a dotted line. The approximate position of \EF\ of SnTe is highlighted with a blue dashed line. The arrow indicates the effect of In doping on \EF\ in \SIT.}
\label{fig2}
\end{figure}

Figure~\ref{fig2}(a) presents the superconducting phase diagram of \SIT\ as determined from resistivity (zero resistance), specific heat [onset of the jump-like anomaly in $\cel(T)$], and theoretically calculated \Tc\ values. Green data points are taken from literature (Refs.~\onlinecite{novak13a} and \onlinecite{haldolaarachchige16a}) for comparison. The experimental \Tc\ values exhibit a dome-like $x$ dependence with a sharp dip-like anomaly centered at $x=0.58$. 

In Fig.~\ref{fig2}(b) the density of states (DOS) is shown against the In concentration $x$. For $x =0$, we find experimentally a sizeable DOS due to the unintentionally doped Sn vacancies giving rise to free charge carriers in otherwise semiconducting SnTe. Upon doping, the experimental DOS increases, exhibits a slight suppression around $x =0.58$, and a maximum around $x=0.7$. Towards InTe, the DOS decreases again. 

Figure~\ref{fig2}(d) shows the calculated DOS for InTe as a function of energy; the Fermi energy \EF\ is taken as the origin (dotted vertical line). The approximate position of SnTe is indicated by a vertical dashed blue line in the rigid-band picture, showing the narrow-gap feature of SnTe. The arrow sketches the effect of In doping, i.e., the band-filling change. Our calculations yield a sharp peak-like anomaly above the Fermi level in InTe which is a van Hove singularity typically found for the rock-salt fcc structure. The effect of the van Hove singularity can also be traced on theoretical results for the DOS as function of $x$ shown in Fig.~\ref{fig2}(b). To readily compare our calculations with the experimental results, the theoretical DOS was corrected for the electron-phonon-interaction-induced enhancement of DOS data estimated from specific-heat measurements, cf.\ Section~S6 in the Supplemental Material \cite{Suppl}. For $x =0$, the calculated DOS is almost zero as expected for an insulator / semiconductor. Upon doping, the DOS increases and exhibits a maximum around $x=0.7$. Towards InTe, the DOS decreases. While the slight suppression in the experimental DOS around $x=0.58$ is not seen in our calculated data, all other features are well reproduced and there is a reasonable agreement between experimental and calculated DOS. 

In Fig.~\ref{fig2}(c) the calculated Eliashberg function $\alpha^2 F$ is plotted against the phonon frequency $\omega$ for the end compound InTe. The integration of $\alpha^2 F$ yields the electron-phonon coupling strength $\lambda$ which is plotted in red. The inset compares the $x$ dependence of theoretical (for $\omega\rightarrow\infty$) and experimental values of $\lambda$. Again, there is a reasonable agreement between experiment and theory except the sizeable enhancement in the calculated data around $x=0.7$.

In the phase diagram in Fig.~\ref{fig2}(a), we also show calculated \Tc\ values. At low doping the calculations qualitatively reproduce the overall tendency of increasing \Tc\ values with $x$ although the absolute values are not matching well. We note that spin-orbit interaction which is not included to the present calculations may account for at least a part of the discrepancy \cite{heid10a}. The maximum in \Tc\ is found around $x=0.7$ which indicates the doping concentration where \EF\ falls onto the van Hove singularity. Although the  maximum \Tc\ value is overestimated in our DFT calculations, the quantitative agreement between experiment and theory is very good above $x \sim 0.7$. 

\begin{figure}[t]
\centering
\includegraphics[width=8.5cm,clip]{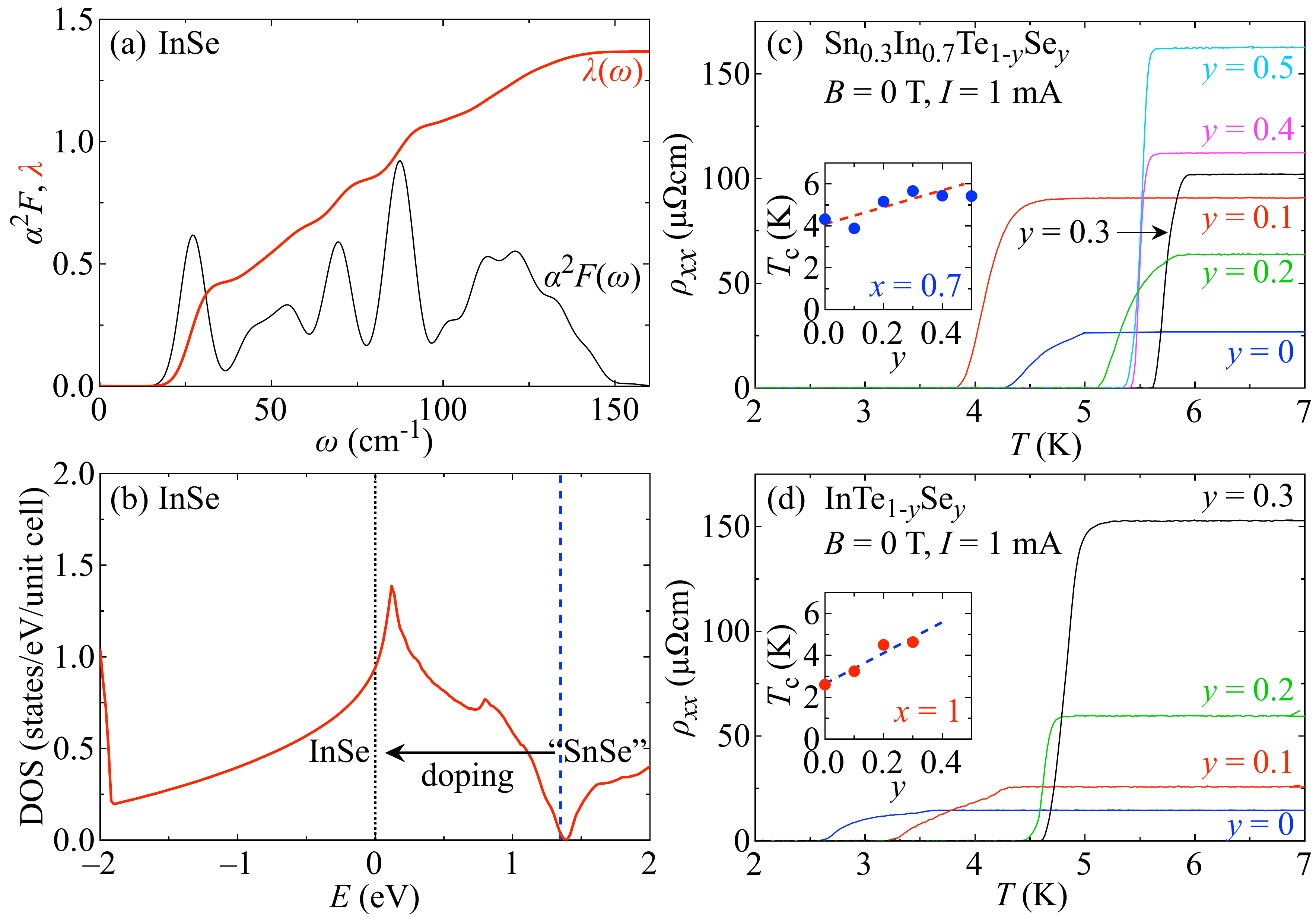}
\caption{(a) Calculated Eliashberg function $\alpha^2 F$ (black) and (integrated) electron-phonon coupling constant $\lambda$ (red) as a function of the phonon frequency $\omega$ for cubic InSe. (b) Calculated DOS for InSe as a function of energy. The Fermi energy \EF\ of InSe is defined as 0 and indicated with a dotted line. The approximate position of \EF\ of hypothetical ``cubic SnSe'' is highlighted with a blue dashed line. The arrow indicates the effect of In doping on \EF\ in ``cubic Sn$_{1-x}$In$_x$Se''. (c) Resistivity data of codoped \SITS\ for fixed $x = 0.7$ and (d) $x = 1$. Insets in both panels show the $y$ dependence of \Tc.}
\label{fig3}
\end{figure}
We also performed DFT calculations for cubic InSe (reported to exist when grown at $\approx 10.4$~GPa \cite{schwarz07a}) to see whether Se codoping on the Te site can lead to a further enhancement of \Tc\ since lighter elements may generally yield higher phonon frequencies and hence higher \Tc\ values. The results are shown in Figs.~\ref{fig3}(a) and \ref{fig3}(b) which contain the same information for InSe as Figs.~\ref{fig2}c and d for InTe. Apparently, the integrated Eliashberg function shown in Fig.~\ref{fig3}(a) yields a $\sim 2.5$ times larger electron-phonon coupling constant $\lambda$ which may give rise to an increased pairing interaction. As shown in Fig.~\ref{fig3}(b), the DOS of InSe exhibits a similar van Hove singularity as found in InTe. In InSe, the singularity lies closer to the Fermi level than in the case of InTe and explains why $\lambda$ is larger in InSe for which the present calculations predict $\Tc=8.5$~K. This suggests the experimental exploration at higher In concentrations for enhanced \Tc\ values by codoping Se. 

Motivated by these DFT calculation results, we attempted to grow \SITS\ crystals. This turned out to be possible up to $y=0.5$ for $x= 0.5 - 0.7$ which is the solubility limit considering the applicable pressure range up to 8~GPa in our high-pressure apparatus. Resistivity data $\rho_{xx}$ for fixed $x =0.7$ and $x = 1$ are shown in Figs.~\ref{fig3}(c) and \ref{fig3}(d), respectively. Although the absolute values of the residual resistivities $\rho_{\rm 6K}$ systematically increase with $y$, all samples exhibit a drop to zero resistivity. The increase in $\rho_{\rm 6K}$ is likely a consequence of higher disorder in these samples due to the introduction of another dopant Se with different ionic size. Nevertheless, as suggested by our DFT calculations, \Tc\ is further enhanced. The inset in each panel shows \Tc\ vs the Se concentration $y$. The strongest enhancement was found for $x=0.9$ and 1 where \Tc\ increases from $\sim 2.6$~K for $y = 0$ to 4.6~K and 5.0~K, respectively, for $y = 0.3$ which is the solubility limit for these high In concentrations. In the case of $x=0.7$, \Tc\ increased from 2.6~K for $y = 0$ to 5.7~K for $y = 0.3$. The latter is the highest \Tc\ found in this study. As can be seen in the inset of Fig.~\ref{fig3}(c), the solubility limit for $x=0.7$ is $y=0.5$ but for this composition \Tc\ tends to slightly decrease again and saturate for higher Se concentrations. The $y$ dependence of other In concentrations $x$ can be found in Fig.~S4 in the Supplemental Material \cite{Suppl}. There we also show in Fig.~S2 the Se-codoping effect on the cubic lattice constant $a_c$ for $x=0.7$ and $x=1$. Due to the smaller ionic radius of Se, $a_c$ shrinks. Another way to compress the lattice is to apply physical pressure $p$. We probed this in the case of InTe: \Tc\ was found to decrease linearly as a function of $p$, see Fig.~S5 in the Supplemental Material \cite{Suppl}. Such a behavior is often seen in conventional superconductors and hence the \Tc\ enhancement by Se codoping is not due to the chemical pressure effect on the crystal lattice. This is in accord with our DFT calculations that the different character of the wave functions when changing from $5p$ (Te) to $4p$ (Se) has a distinct effect on the pairing interaction.

\begin{figure}[t]
\centering
\includegraphics[width=8.5cm,clip]{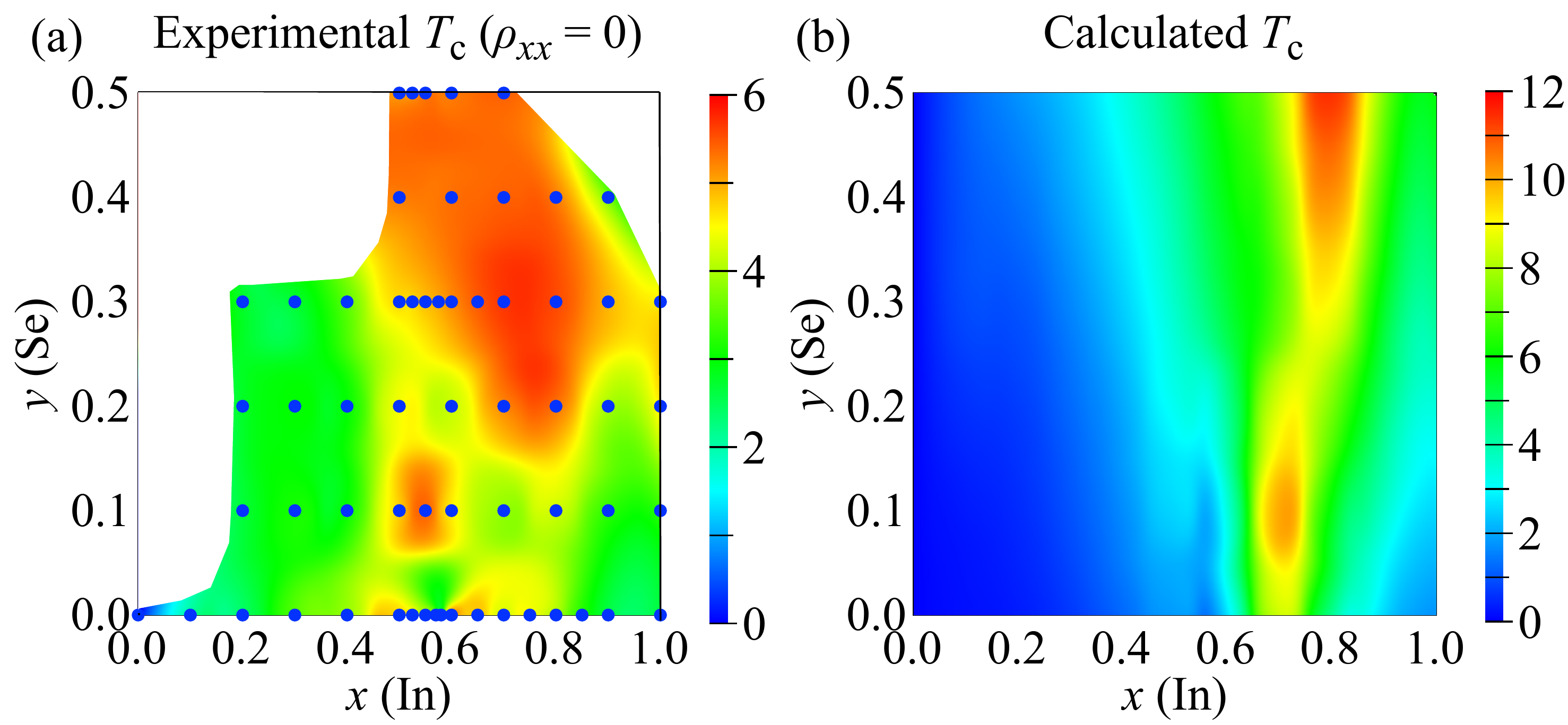}
\caption{(a) Experimentally and (b) theoretically determined superconducting phase diagram of \SITS\ as functions of $x$ and $y$. The blue symbols in (a) indicate the samples $(x,y)$, for which \Tc\ was actually measured. The white areas were not explored and are partially beyond the solubility limit for alloying \SITS. It should be noted that the scale in (a) and (b) differs by a factor of 2, therefore the color scheme is not the same in both panels.}
\label{fig4}
\end{figure}
Figures~\ref{fig4}(a) and \ref{fig4}(b) provide a comparison between measured and calculated \Tc\ values of \SITS\ as functions of $x$ and $y$. At low $x$, the DFT calculation systematically underestimates \Tc. This is perhaps due to the rigid-band approximation and employing it to InTe and InSe rather than SnTe and hypothetic ``cubic SnSe'', respectively. The real band structure may change upon doping beyond the rigid band approximation. Nevertheless, the tendency towards enhanced \Tc\ values around $x\geq 0.6$ and $y\geq 0.3$ is correctly reproduced and one can safely conclude that the optimal $x$ of the superconducting dome shifts towards $x=1$ with increasing $y$.

Finally, we discuss a possible scenario which can explain the observed features. Apparently, In and Se codoping into SnTe have the capability to increase \Tc\ from $<0.3$~K up to almost 6~K -- or possibly even more with higher Se content. One scenario which attracted considerable interest in literature is the ``negative-$U$ mechanism'' which relies on valence-skipping elements \cite{varma88a,dzero05a,hase16a}. Nominally In should replace Sn in an isovalent manner. However, the formal In$^{2+}$ state can be energetically unstable. In is then expected to appear as In$^{1+}$ ($4d^{10}5s^2$), In$^{3+}$ ($4d^{10}5s^0$) or even a mixture of them. Depending on the band filling, this may lead to, e.g., diamagnetic insulating or metallic behavior, a charge-Kondo effect, or possibly enhanced superconductivity \cite{varma88a,dzero05a,hase16a}. Moreover, when the valence-skipping states order, a charge-density wave (CDW) can be expected. The ``negative-$U$ mechanism'' is, for example, considered to be responsible for the observed strong enhancements of \Tc\ in \PTT\ and doped BaBiO$_3$ \cite{matsushita05a,cava88a}. There is indeed support for the assumption that the In valence state plays a significant role in \SIT: A slope change in $\Tc(x)$ was reported for $x \sim 0.08 - 0.1$ \cite{haldolaarachchige16a}, which coincides with a change from hole-doping (i.e., In$^{1+}$) to electron-doping (i.e., In$^{3+}$), and also with a slope change in the evolution of the cubic lattice constant when crossing $x \approx 0.1$, explainable with a change in the In valence states. Based on DFT calculations, it was also proposed that an In impurity band forms in \SIT\ which intersects the Fermi energy and consists of hybridized In-$5s$ and Te-$5p$ states \cite{haldolaarachchige16a}.
In such a scenario, it is also possible to understand phenomenologically the sharp suppression of \Tc\ and superconducting volume fraction around $x = 0.58$ in \SIT. Such a dip structure of the superconducting phase diagram is sometimes encountered in unconventional systems, such as high-\Tc\ cuprates or iron pnictides. The doping concentration range where the superconductivity is suppressed is usually close to the onset of different orders and competing phases (e.g., stripe order (Ref.~\onlinecite{tranquada12a}) for $x\sim 0.125$ in La$_{2-x}$Ba$_{x}$CuO$_4$ or structural and magnetic order (Ref.~\cite{hiraishi14a}) for $x\sim 0.2$ in LaFeAsO$_{1-x}$H$_x$). One may speculate that in \SIT\ (and \SITS) a certain In$^{1+}$--In$^{3+}$ order forms out, e.g., a CDW supported by the apparent Fermi surface instability at $x = 0.58$ and competes with the superconductivity. There might even be a critical $x$ value for which the superconductivity is completely suppressed. The very good quantitative agreement between experimental and calculated \Tc\ values in the highly-doped region of the phase diagram [Fig.~\ref{fig2}(a)] could then indicate that the negative-$U$ mechanism is not at work any more for $x \geq 0.8$ and the system (for $y=0$) is simply metallic with nominal In$^{2+}$ valence state forming a conventional BCS superconductor at low temperatures as it is also supported by the physical-pressure effect on \Tc. However, we could not find any experimental evidence yet for a CDW formation in \SITS\ which could be a promising starting point for future works. 

\nocite{kriener08a,kriener11a,tinkham96,mcmillan68a,muehlschlegel59a,padamsee73a,perdew96a,vanderbilt90a,mostofi14a,koretsune17a}
 

\begin{thebibliography}{44}
\expandafter\ifx\csname natexlab\endcsname\relax\def\natexlab#1{#1}\fi
\expandafter\ifx\csname bibnamefont\endcsname\relax
  \def\bibnamefont#1{#1}\fi
\expandafter\ifx\csname bibfnamefont\endcsname\relax
  \def\bibfnamefont#1{#1}\fi
\expandafter\ifx\csname citenamefont\endcsname\relax
  \def\citenamefont#1{#1}\fi
\expandafter\ifx\csname url\endcsname\relax
  \def\url#1{\texttt{#1}}\fi
\expandafter\ifx\csname urlprefix\endcsname\relax\def\urlprefix{URL }\fi
\providecommand{\bibinfo}[2]{#2}
\providecommand{\eprint}[2][]{\url{#2}}

\bibitem[{\citenamefont{Hasan and Kane}(2010)}]{hasan10a}
\bibinfo{author}{\bibfnamefont{M.}~\bibnamefont{Hasan}} \bibnamefont{and}
  \bibinfo{author}{\bibfnamefont{C.}~\bibnamefont{Kane}},
  \bibinfo{journal}{Rev.\ Mod.\ Phys.} \textbf{\bibinfo{volume}{82}},
  \bibinfo{pages}{3045} (\bibinfo{year}{2010}).

\bibitem[{\citenamefont{Qi and Zhang}(2011)}]{qi11a}
\bibinfo{author}{\bibfnamefont{X.-L.} \bibnamefont{Qi}} \bibnamefont{and}
  \bibinfo{author}{\bibfnamefont{S.-C.} \bibnamefont{Zhang}},
  \bibinfo{journal}{Rev.\ Mod.\ Phys.} \textbf{\bibinfo{volume}{83}},
  \bibinfo{pages}{1057} (\bibinfo{year}{2011}).

\bibitem[{\citenamefont{Ando and Fu}(2015)}]{ando15a}
\bibinfo{author}{\bibfnamefont{Y.}~\bibnamefont{Ando}} \bibnamefont{and}
  \bibinfo{author}{\bibfnamefont{L.}~\bibnamefont{Fu}}, \bibinfo{journal}{Ann.\
  Rev.\ Cond.\ Matter Phys.} \textbf{\bibinfo{volume}{6}}, \bibinfo{pages}{361}
  (\bibinfo{year}{2015}).

\bibitem[{\citenamefont{Hsieh et~al.}(2012)\citenamefont{Hsieh, Lin, Liu, Duan,
  Bansil, and L.Fu}}]{hsieh12b}
\bibinfo{author}{\bibfnamefont{T.~H.} \bibnamefont{Hsieh}},
  \bibinfo{author}{\bibfnamefont{H.}~\bibnamefont{Lin}},
  \bibinfo{author}{\bibfnamefont{J.}~\bibnamefont{Liu}},
  \bibinfo{author}{\bibfnamefont{W.}~\bibnamefont{Duan}},
  \bibinfo{author}{\bibfnamefont{A.}~\bibnamefont{Bansil}}, \bibnamefont{and}
  \bibinfo{author}{\bibnamefont{L.Fu}}, \bibinfo{journal}{Nat.\ Commun.}
  \textbf{\bibinfo{volume}{3}}, \bibinfo{pages}{982} (\bibinfo{year}{2012}).

\bibitem[{\citenamefont{Tanaka et~al.}(2012)\citenamefont{Tanaka, Ren, Sato,
  Nakayama, Souma, Takahashi, Segawa, and Ando}}]{ytanaka12b}
\bibinfo{author}{\bibfnamefont{Y.}~\bibnamefont{Tanaka}},
  \bibinfo{author}{\bibfnamefont{Z.}~\bibnamefont{Ren}},
  \bibinfo{author}{\bibfnamefont{T.}~\bibnamefont{Sato}},
  \bibinfo{author}{\bibfnamefont{K.}~\bibnamefont{Nakayama}},
  \bibinfo{author}{\bibfnamefont{S.}~\bibnamefont{Souma}},
  \bibinfo{author}{\bibfnamefont{T.}~\bibnamefont{Takahashi}},
  \bibinfo{author}{\bibfnamefont{K.}~\bibnamefont{Segawa}}, \bibnamefont{and}
  \bibinfo{author}{\bibfnamefont{Y.}~\bibnamefont{Ando}},
  \bibinfo{journal}{Nature Phys.} \textbf{\bibinfo{volume}{8}},
  \bibinfo{pages}{800} (\bibinfo{year}{2012}).

\bibitem[{\citenamefont{Fu}(2011)}]{fu11a}
\bibinfo{author}{\bibfnamefont{L.}~\bibnamefont{Fu}}, \bibinfo{journal}{Phys.\
  Rev.\ Lett.} \textbf{\bibinfo{volume}{106}}, \bibinfo{pages}{106802}
  (\bibinfo{year}{2011}).

\bibitem[{\citenamefont{Hein et~al.}(1965)\citenamefont{Hein, Gibson,
  S.Ãllgaier, B.~B.~Housteon, Mazelsky, and Miller}}]{hein65a}
\bibinfo{author}{\bibfnamefont{R.~A.} \bibnamefont{Hein}},
  \bibinfo{author}{\bibfnamefont{J.~W.} \bibnamefont{Gibson}},
  \bibinfo{author}{\bibfnamefont{R.}~\bibnamefont{S.Ãllgaier}},
  \bibinfo{author}{\bibfnamefont{J.}~\bibnamefont{B.~B.~Housteon}},
  \bibinfo{author}{\bibfnamefont{R.}~\bibnamefont{Mazelsky}}, \bibnamefont{and}
  \bibinfo{author}{\bibfnamefont{R.~C.} \bibnamefont{Miller}},
  \bibinfo{journal}{Proc.\ of the 9$^{\rm th}$ Int.\ Conf.\ on Low Temp.\
  Phys.} p. \bibinfo{pages}{604} (\bibinfo{year}{1965}).

\bibitem[{\citenamefont{Nakajima et~al.}(1973)\citenamefont{Nakajima, Isino,
  Miyauchi, and Kanda}}]{nakajima73a}
\bibinfo{author}{\bibfnamefont{T.}~\bibnamefont{Nakajima}},
  \bibinfo{author}{\bibfnamefont{M.}~\bibnamefont{Isino}},
  \bibinfo{author}{\bibfnamefont{H.}~\bibnamefont{Miyauchi}}, \bibnamefont{and}
  \bibinfo{author}{\bibfnamefont{E.}~\bibnamefont{Kanda}},
  \bibinfo{journal}{J.\ Phys.\ Soc.\ Jpn.} \textbf{\bibinfo{volume}{34}},
  \bibinfo{pages}{282} (\bibinfo{year}{1973}).

\bibitem[{\citenamefont{Bushmarina et~al.}(1985)\citenamefont{Bushmarina,
  Drabkin, Kompaniets, Parfen'ev, Shamshur, and Shakhov}}]{bushmarina86a}
\bibinfo{author}{\bibfnamefont{G.~S.} \bibnamefont{Bushmarina}},
  \bibinfo{author}{\bibfnamefont{I.~A.} \bibnamefont{Drabkin}},
  \bibinfo{author}{\bibfnamefont{V.~V.} \bibnamefont{Kompaniets}},
  \bibinfo{author}{\bibfnamefont{R.~V.} \bibnamefont{Parfen'ev}},
  \bibinfo{author}{\bibfnamefont{D.~V.} \bibnamefont{Shamshur}},
  \bibnamefont{and} \bibinfo{author}{\bibfnamefont{M.~A.}
  \bibnamefont{Shakhov}}, \bibinfo{journal}{Sov.\ Phys.\ Solid State}
  \textbf{\bibinfo{volume}{28}}, \bibinfo{pages}{612} (\bibinfo{year}{1985}).

\bibitem[{\citenamefont{Erickson et~al.}(2009)\citenamefont{Erickson, Chu,
  Toney, Geballe, and Fisher}}]{erickson09a}
\bibinfo{author}{\bibfnamefont{A.~S.} \bibnamefont{Erickson}},
  \bibinfo{author}{\bibfnamefont{J.-H.} \bibnamefont{Chu}},
  \bibinfo{author}{\bibfnamefont{M.~F.} \bibnamefont{Toney}},
  \bibinfo{author}{\bibfnamefont{T.~H.} \bibnamefont{Geballe}},
  \bibnamefont{and} \bibinfo{author}{\bibfnamefont{I.~R.}
  \bibnamefont{Fisher}}, \bibinfo{journal}{Phys.\ Rev.\ B}
  \textbf{\bibinfo{volume}{79}}, \bibinfo{pages}{024520}
  (\bibinfo{year}{2009}).

\bibitem[{\citenamefont{Sasaki et~al.}(2012)\citenamefont{Sasaki, Ren, Taskin,
  Segawa, Fu, and Ando}}]{sasaki12a}
\bibinfo{author}{\bibfnamefont{S.}~\bibnamefont{Sasaki}},
  \bibinfo{author}{\bibfnamefont{Z.}~\bibnamefont{Ren}},
  \bibinfo{author}{\bibfnamefont{A.~A.} \bibnamefont{Taskin}},
  \bibinfo{author}{\bibfnamefont{K.}~\bibnamefont{Segawa}},
  \bibinfo{author}{\bibfnamefont{L.}~\bibnamefont{Fu}}, \bibnamefont{and}
  \bibinfo{author}{\bibfnamefont{Y.}~\bibnamefont{Ando}},
  \bibinfo{journal}{Phys.\ Rev.\ Lett.} \textbf{\bibinfo{volume}{109}},
  \bibinfo{pages}{217004} (\bibinfo{year}{2012}).

\bibitem[{\citenamefont{Sato et~al.}(2013)\citenamefont{Sato, Tanaka, Nakayama,
  Souma, Takahashi, Sasaki, Ren, Taskin, Segawa, and Ando}}]{tsato13a}
\bibinfo{author}{\bibfnamefont{T.}~\bibnamefont{Sato}},
  \bibinfo{author}{\bibfnamefont{Y.}~\bibnamefont{Tanaka}},
  \bibinfo{author}{\bibfnamefont{K.}~\bibnamefont{Nakayama}},
  \bibinfo{author}{\bibfnamefont{S.}~\bibnamefont{Souma}},
  \bibinfo{author}{\bibfnamefont{T.}~\bibnamefont{Takahashi}},
  \bibinfo{author}{\bibfnamefont{S.}~\bibnamefont{Sasaki}},
  \bibinfo{author}{\bibfnamefont{Z.}~\bibnamefont{Ren}},
  \bibinfo{author}{\bibfnamefont{A.~A.} \bibnamefont{Taskin}},
  \bibinfo{author}{\bibfnamefont{K.}~\bibnamefont{Segawa}}, \bibnamefont{and}
  \bibinfo{author}{\bibfnamefont{Y.}~\bibnamefont{Ando}},
  \bibinfo{journal}{Phys.\ Rev.\ Lett.} \textbf{\bibinfo{volume}{110}},
  \bibinfo{pages}{206804} (\bibinfo{year}{2013}).

\bibitem[{\citenamefont{Novak et~al.}(2013)\citenamefont{Novak, Sasaki,
  Kriener, Segawa, and Ando}}]{novak13a}
\bibinfo{author}{\bibfnamefont{M.}~\bibnamefont{Novak}},
  \bibinfo{author}{\bibfnamefont{S.}~\bibnamefont{Sasaki}},
  \bibinfo{author}{\bibfnamefont{M.}~\bibnamefont{Kriener}},
  \bibinfo{author}{\bibfnamefont{K.}~\bibnamefont{Segawa}}, \bibnamefont{and}
  \bibinfo{author}{\bibfnamefont{Y.}~\bibnamefont{Ando}},
  \bibinfo{journal}{Phys.\ Rev.\ B} \textbf{\bibinfo{volume}{88}},
  \bibinfo{pages}{140502(R)} (\bibinfo{year}{2013}).

\bibitem[{\citenamefont{Zhong et~al.}(2013)\citenamefont{Zhong, Schneeloch,
  Shi, Xu, Zhang, Tranquada, Li, and Gu}}]{rdzhong13a}
\bibinfo{author}{\bibfnamefont{R.~D.} \bibnamefont{Zhong}},
  \bibinfo{author}{\bibfnamefont{J.~A.} \bibnamefont{Schneeloch}},
  \bibinfo{author}{\bibfnamefont{X.~Y.} \bibnamefont{Shi}},
  \bibinfo{author}{\bibfnamefont{Z.~J.} \bibnamefont{Xu}},
  \bibinfo{author}{\bibfnamefont{C.}~\bibnamefont{Zhang}},
  \bibinfo{author}{\bibfnamefont{J.~M.} \bibnamefont{Tranquada}},
  \bibinfo{author}{\bibfnamefont{Q.}~\bibnamefont{Li}}, \bibnamefont{and}
  \bibinfo{author}{\bibfnamefont{G.~D.} \bibnamefont{Gu}},
  \bibinfo{journal}{Phys.\ Rev.\ B} \textbf{\bibinfo{volume}{88}},
  \bibinfo{pages}{020505(R)} (\bibinfo{year}{2013}).

\bibitem[{\citenamefont{He et~al.}(2013)\citenamefont{He, Zhang, Pan, Hong,
  Zhou, and Li}}]{lphe13a}
\bibinfo{author}{\bibfnamefont{L.~P.} \bibnamefont{He}},
  \bibinfo{author}{\bibfnamefont{Z.}~\bibnamefont{Zhang}},
  \bibinfo{author}{\bibfnamefont{J.}~\bibnamefont{Pan}},
  \bibinfo{author}{\bibfnamefont{X.~C.} \bibnamefont{Hong}},
  \bibinfo{author}{\bibfnamefont{S.~Y.} \bibnamefont{Zhou}}, \bibnamefont{and}
  \bibinfo{author}{\bibfnamefont{S.~Y.} \bibnamefont{Li}},
  \bibinfo{journal}{Phys.\ Rev.\ B} \textbf{\bibinfo{volume}{88}},
  \bibinfo{pages}{014523} (\bibinfo{year}{2013}).

\bibitem[{\citenamefont{Maurya et~al.}(2014)\citenamefont{Maurya, andP.
  Shrivastava, and Patnaik}}]{maurya14a}
\bibinfo{author}{\bibfnamefont{V.~K.} \bibnamefont{Maurya}},
  \bibinfo{author}{\bibfnamefont{S.}~\bibnamefont{andP. Shrivastava}},
  \bibnamefont{and} \bibinfo{author}{\bibfnamefont{S.}~\bibnamefont{Patnaik}},
  \bibinfo{journal}{Europhys.\ Lett.} \textbf{\bibinfo{volume}{108}},
  \bibinfo{pages}{37010} (\bibinfo{year}{2014}).

\bibitem[{\citenamefont{Saghir et~al.}(2014)\citenamefont{Saghir, Barker,
  Balakrishnan, Hillier, and Lees}}]{saghir14a}
\bibinfo{author}{\bibfnamefont{M.}~\bibnamefont{Saghir}},
  \bibinfo{author}{\bibfnamefont{J.~A.~T.} \bibnamefont{Barker}},
  \bibinfo{author}{\bibfnamefont{G.}~\bibnamefont{Balakrishnan}},
  \bibinfo{author}{\bibfnamefont{A.~D.} \bibnamefont{Hillier}},
  \bibnamefont{and} \bibinfo{author}{\bibfnamefont{M.~R.} \bibnamefont{Lees}},
  \bibinfo{journal}{Phys.\ Rev.\ B} \textbf{\bibinfo{volume}{90}},
  \bibinfo{pages}{064508} (\bibinfo{year}{2014}).

\bibitem[{\citenamefont{Hashimoto et~al.}(2015)\citenamefont{Hashimoto, Yada,
  Sato, and Tanaka}}]{hashimoto15a}
\bibinfo{author}{\bibfnamefont{T.}~\bibnamefont{Hashimoto}},
  \bibinfo{author}{\bibfnamefont{K.}~\bibnamefont{Yada}},
  \bibinfo{author}{\bibfnamefont{M.}~\bibnamefont{Sato}}, \bibnamefont{and}
  \bibinfo{author}{\bibfnamefont{Y.}~\bibnamefont{Tanaka}},
  \bibinfo{journal}{Phys.\ Rev.\ B} \textbf{\bibinfo{volume}{92}},
  \bibinfo{pages}{174527} (\bibinfo{year}{2015}).

\bibitem[{\citenamefont{Haldolaarachchige
  et~al.}(2016)\citenamefont{Haldolaarachchige, Gibson, Xie, Nielsen, Kushwaha,
  and Cava}}]{haldolaarachchige16a}
\bibinfo{author}{\bibfnamefont{N.}~\bibnamefont{Haldolaarachchige}},
  \bibinfo{author}{\bibfnamefont{Q.}~\bibnamefont{Gibson}},
  \bibinfo{author}{\bibfnamefont{W.}~\bibnamefont{Xie}},
  \bibinfo{author}{\bibfnamefont{M.~B.} \bibnamefont{Nielsen}},
  \bibinfo{author}{\bibfnamefont{S.}~\bibnamefont{Kushwaha}}, \bibnamefont{and}
  \bibinfo{author}{\bibfnamefont{R.~J.} \bibnamefont{Cava}},
  \bibinfo{journal}{Phys.\ Rev.\ B} \textbf{\bibinfo{volume}{93}},
  \bibinfo{pages}{024520} (\bibinfo{year}{2016}).

\bibitem[{\citenamefont{Maeda et~al.}(2017)\citenamefont{Maeda, Hirose, Matano,
  Novak, Ando, and q.~Zheng}}]{maeda17b}
\bibinfo{author}{\bibfnamefont{S.}~\bibnamefont{Maeda}},
  \bibinfo{author}{\bibfnamefont{R.}~\bibnamefont{Hirose}},
  \bibinfo{author}{\bibfnamefont{K.}~\bibnamefont{Matano}},
  \bibinfo{author}{\bibfnamefont{M.}~\bibnamefont{Novak}},
  \bibinfo{author}{\bibfnamefont{Y.}~\bibnamefont{Ando}}, \bibnamefont{and}
  \bibinfo{author}{\bibfnamefont{G.}~\bibnamefont{q.~Zheng}},
  \bibinfo{journal}{Phys.\ Rev.\ B} \textbf{\bibinfo{volume}{96}},
  \bibinfo{pages}{104052} (\bibinfo{year}{2017}).

\bibitem[{\citenamefont{B\"{o}mmel et~al.}(1963)\citenamefont{B\"{o}mmel,
  Darnell, Libby, Tittmann, and Yencha}}]{boemmel63a}
\bibinfo{author}{\bibfnamefont{H.~E.} \bibnamefont{B\"{o}mmel}},
  \bibinfo{author}{\bibfnamefont{A.~J.} \bibnamefont{Darnell}},
  \bibinfo{author}{\bibfnamefont{W.~F.} \bibnamefont{Libby}},
  \bibinfo{author}{\bibfnamefont{B.~R.} \bibnamefont{Tittmann}},
  \bibnamefont{and} \bibinfo{author}{\bibfnamefont{A.~J.}
  \bibnamefont{Yencha}}, \bibinfo{journal}{Science}
  \textbf{\bibinfo{volume}{141}}, \bibinfo{pages}{714} (\bibinfo{year}{1963}).

\bibitem[{\citenamefont{Darnell et~al.}(1963)\citenamefont{Darnell, Yencha, and
  Libby}}]{darnell63a}
\bibinfo{author}{\bibfnamefont{A.~J.} \bibnamefont{Darnell}},
  \bibinfo{author}{\bibfnamefont{A.~J.} \bibnamefont{Yencha}},
  \bibnamefont{and} \bibinfo{author}{\bibfnamefont{W.~F.} \bibnamefont{Libby}},
  \bibinfo{journal}{Science} \textbf{\bibinfo{volume}{141}},
  \bibinfo{pages}{713} (\bibinfo{year}{1963}).

\bibitem[{Sup()}]{Suppl}
\bibinfo{howpublished}{{See Supplemental Material at [URL will be inserted by
  publisher] for complementing data.}}

\bibitem[{\citenamefont{Giannozzi et~al.}(2009)\citenamefont{Giannozzi, Baroni,
  Bonini, Calandra, Car, Cavazzoni, Ceresoli, Chiarotti, Cococcioni, Dabo1
  et~al.}}]{giannozzi09a}
\bibinfo{author}{\bibfnamefont{P.}~\bibnamefont{Giannozzi}},
  \bibinfo{author}{\bibfnamefont{S.}~\bibnamefont{Baroni}},
  \bibinfo{author}{\bibfnamefont{N.}~\bibnamefont{Bonini}},
  \bibinfo{author}{\bibfnamefont{M.}~\bibnamefont{Calandra}},
  \bibinfo{author}{\bibfnamefont{R.}~\bibnamefont{Car}},
  \bibinfo{author}{\bibfnamefont{C.}~\bibnamefont{Cavazzoni}},
  \bibinfo{author}{\bibfnamefont{D.}~\bibnamefont{Ceresoli}},
  \bibinfo{author}{\bibfnamefont{G.~L.} \bibnamefont{Chiarotti}},
  \bibinfo{author}{\bibfnamefont{M.}~\bibnamefont{Cococcioni}},
  \bibinfo{author}{\bibfnamefont{I.}~\bibnamefont{Dabo1}},
  \bibnamefont{et~al.}, \bibinfo{journal}{J.\ Phys.: Condens.\ Matter}
  \textbf{\bibinfo{volume}{21}}, \bibinfo{pages}{395502}
  (\bibinfo{year}{2009}).

\bibitem[{\citenamefont{Allen and Dynes}(1975)}]{allen75a}
\bibinfo{author}{\bibfnamefont{P.~B.} \bibnamefont{Allen}} \bibnamefont{and}
  \bibinfo{author}{\bibfnamefont{R.~C.} \bibnamefont{Dynes}},
  \bibinfo{journal}{Phys.\ Rev.\ B} \textbf{\bibinfo{volume}{12}},
  \bibinfo{pages}{905} (\bibinfo{year}{1975}).

\bibitem[{\citenamefont{Heid et~al.}(2010)\citenamefont{Heid, Bohnen,
  Sklyadneva, and Chulkov}}]{heid10a}
\bibinfo{author}{\bibfnamefont{R.}~\bibnamefont{Heid}},
  \bibinfo{author}{\bibfnamefont{K.-P.} \bibnamefont{Bohnen}},
  \bibinfo{author}{\bibfnamefont{I.~Y.} \bibnamefont{Sklyadneva}},
  \bibnamefont{and} \bibinfo{author}{\bibfnamefont{E.~V.}
  \bibnamefont{Chulkov}}, \bibinfo{journal}{Phys.\ Rev.\ B}
  \textbf{\bibinfo{volume}{81}}, \bibinfo{pages}{174527}
  (\bibinfo{year}{2010}).

\bibitem[{\citenamefont{Schwarz et~al.}(2007)\citenamefont{Schwarz, Go{\~{n}}i,
  Syassen, Cantarero, and Chevy}}]{schwarz07a}
\bibinfo{author}{\bibfnamefont{U.}~\bibnamefont{Schwarz}},
  \bibinfo{author}{\bibfnamefont{A.~R.} \bibnamefont{Go{\~{n}}i}},
  \bibinfo{author}{\bibfnamefont{K.}~\bibnamefont{Syassen}},
  \bibinfo{author}{\bibfnamefont{A.}~\bibnamefont{Cantarero}},
  \bibnamefont{and} \bibinfo{author}{\bibfnamefont{A.}~\bibnamefont{Chevy}},
  \bibinfo{journal}{High Pressure Research} \textbf{\bibinfo{volume}{8}},
  \bibinfo{pages}{396} (\bibinfo{year}{2007}).

\bibitem[{\citenamefont{Varma}(1988)}]{varma88a}
\bibinfo{author}{\bibfnamefont{C.}~\bibnamefont{Varma}},
  \bibinfo{journal}{Phys.\ Rev.\ Lett.} \textbf{\bibinfo{volume}{61}},
  \bibinfo{pages}{2713} (\bibinfo{year}{1988}).

\bibitem[{\citenamefont{Dzero and Schmalian}(2005)}]{dzero05a}
\bibinfo{author}{\bibfnamefont{M.}~\bibnamefont{Dzero}} \bibnamefont{and}
  \bibinfo{author}{\bibfnamefont{J.}~\bibnamefont{Schmalian}},
  \bibinfo{journal}{Phys.\ Rev.\ Lett.} \textbf{\bibinfo{volume}{94}},
  \bibinfo{pages}{157003} (\bibinfo{year}{2005}).

\bibitem[{\citenamefont{Hase et~al.}(2016)\citenamefont{Hase, Yasutomi,
  Yanagisawa, Odagiri, and Nishio}}]{hase16a}
\bibinfo{author}{\bibfnamefont{I.}~\bibnamefont{Hase}},
  \bibinfo{author}{\bibfnamefont{K.}~\bibnamefont{Yasutomi}},
  \bibinfo{author}{\bibfnamefont{T.}~\bibnamefont{Yanagisawa}},
  \bibinfo{author}{\bibfnamefont{K.}~\bibnamefont{Odagiri}}, \bibnamefont{and}
  \bibinfo{author}{\bibfnamefont{T.}~\bibnamefont{Nishio}},
  \bibinfo{journal}{Physica C} \textbf{\bibinfo{volume}{527}},
  \bibinfo{pages}{85} (\bibinfo{year}{2016}).

\bibitem[{\citenamefont{Matsushita et~al.}(2005)\citenamefont{Matsushita,
  Bluhm, Geballe, and Fisher}}]{matsushita05a}
\bibinfo{author}{\bibfnamefont{Y.}~\bibnamefont{Matsushita}},
  \bibinfo{author}{\bibfnamefont{H.}~\bibnamefont{Bluhm}},
  \bibinfo{author}{\bibfnamefont{T.~H.} \bibnamefont{Geballe}},
  \bibnamefont{and} \bibinfo{author}{\bibfnamefont{I.~R.}
  \bibnamefont{Fisher}}, \bibinfo{journal}{Phys.\ Rev.\ Lett.}
  \textbf{\bibinfo{volume}{94}}, \bibinfo{pages}{157002}
  (\bibinfo{year}{2005}).

\bibitem[{\citenamefont{Cava et~al.}(1988)\citenamefont{Cava, Batlogg,
  Krajewski, Farrow, Jr, White, Short, Peck, and Kometani}}]{cava88a}
\bibinfo{author}{\bibfnamefont{R.~J.} \bibnamefont{Cava}},
  \bibinfo{author}{\bibfnamefont{B.}~\bibnamefont{Batlogg}},
  \bibinfo{author}{\bibfnamefont{J.~J.} \bibnamefont{Krajewski}},
  \bibinfo{author}{\bibfnamefont{R.}~\bibnamefont{Farrow}},
  \bibinfo{author}{\bibfnamefont{L.~W.~R.} \bibnamefont{Jr}},
  \bibinfo{author}{\bibfnamefont{A.~E.} \bibnamefont{White}},
  \bibinfo{author}{\bibfnamefont{K.}~\bibnamefont{Short}},
  \bibinfo{author}{\bibfnamefont{W.~F.} \bibnamefont{Peck}}, \bibnamefont{and}
  \bibinfo{author}{\bibfnamefont{T.}~\bibnamefont{Kometani}},
  \bibinfo{journal}{Nature (London)} \textbf{\bibinfo{volume}{332}},
  \bibinfo{pages}{814} (\bibinfo{year}{1988}).

\bibitem[{\citenamefont{Tranquada}(2012)}]{tranquada12a}
\bibinfo{author}{\bibfnamefont{J.~M.} \bibnamefont{Tranquada}},
  \bibinfo{journal}{Physica B} \textbf{\bibinfo{volume}{407}},
  \bibinfo{pages}{1771} (\bibinfo{year}{2012}).

\bibitem[{\citenamefont{Hiraishi et~al.}(2014)\citenamefont{Hiraishi, Iimura,
  Kojima, Yamaura, Hiraka, Ikeda, Miao, Ishikawa, Torii, Miyazaki
  et~al.}}]{hiraishi14a}
\bibinfo{author}{\bibfnamefont{M.}~\bibnamefont{Hiraishi}},
  \bibinfo{author}{\bibfnamefont{S.}~\bibnamefont{Iimura}},
  \bibinfo{author}{\bibfnamefont{K.~M.} \bibnamefont{Kojima}},
  \bibinfo{author}{\bibfnamefont{J.}~\bibnamefont{Yamaura}},
  \bibinfo{author}{\bibfnamefont{H.}~\bibnamefont{Hiraka}},
  \bibinfo{author}{\bibfnamefont{K.}~\bibnamefont{Ikeda}},
  \bibinfo{author}{\bibfnamefont{P.}~\bibnamefont{Miao}},
  \bibinfo{author}{\bibfnamefont{Y.}~\bibnamefont{Ishikawa}},
  \bibinfo{author}{\bibfnamefont{S.}~\bibnamefont{Torii}},
  \bibinfo{author}{\bibfnamefont{M.}~\bibnamefont{Miyazaki}},
  \bibnamefont{et~al.}, \bibinfo{journal}{Nature Phys.}
  \textbf{\bibinfo{volume}{10}}, \bibinfo{pages}{300} (\bibinfo{year}{2014}).

\bibitem[{\citenamefont{Kriener et~al.}(2008)\citenamefont{Kriener, Maeno,
  Oguchi, Ren, Kato, Muranaka, and Akimitsu}}]{kriener08a}
\bibinfo{author}{\bibfnamefont{M.}~\bibnamefont{Kriener}},
  \bibinfo{author}{\bibfnamefont{Y.}~\bibnamefont{Maeno}},
  \bibinfo{author}{\bibfnamefont{T.}~\bibnamefont{Oguchi}},
  \bibinfo{author}{\bibfnamefont{Z.}~\bibnamefont{Ren}},
  \bibinfo{author}{\bibfnamefont{J.}~\bibnamefont{Kato}},
  \bibinfo{author}{\bibfnamefont{T.}~\bibnamefont{Muranaka}}, \bibnamefont{and}
  \bibinfo{author}{\bibfnamefont{J.}~\bibnamefont{Akimitsu}},
  \bibinfo{journal}{Phys.\ Rev.\ B} \textbf{\bibinfo{volume}{78}},
  \bibinfo{pages}{024517} (\bibinfo{year}{2008}).

\bibitem[{\citenamefont{Kriener et~al.}(2011)\citenamefont{Kriener, Segawa,
  Ren, Sasaki, and Ando}}]{kriener11a}
\bibinfo{author}{\bibfnamefont{M.}~\bibnamefont{Kriener}},
  \bibinfo{author}{\bibfnamefont{K.}~\bibnamefont{Segawa}},
  \bibinfo{author}{\bibfnamefont{Z.}~\bibnamefont{Ren}},
  \bibinfo{author}{\bibfnamefont{S.}~\bibnamefont{Sasaki}}, \bibnamefont{and}
  \bibinfo{author}{\bibfnamefont{Y.}~\bibnamefont{Ando}},
  \bibinfo{journal}{Phys.\ Rev.\ Lett.} \textbf{\bibinfo{volume}{106}},
  \bibinfo{pages}{127004} (\bibinfo{year}{2011}).

\bibitem[{\citenamefont{Tinkham}(1996)}]{tinkham96}
\bibinfo{author}{\bibfnamefont{M.}~\bibnamefont{Tinkham}},
  \emph{\bibinfo{title}{Introduction to Superconductivity}}
  (\bibinfo{publisher}{McGraw-Hill}, \bibinfo{year}{1996}),
  \bibinfo{edition}{2nd} ed.

\bibitem[{\citenamefont{McMillan}(1968)}]{mcmillan68a}
\bibinfo{author}{\bibfnamefont{W.}~\bibnamefont{McMillan}},
  \bibinfo{journal}{Phys.\ Rev.} \textbf{\bibinfo{volume}{167}},
  \bibinfo{pages}{331} (\bibinfo{year}{1968}).

\bibitem[{\citenamefont{M\"uhlschlegel}(1959)}]{muehlschlegel59a}
\bibinfo{author}{\bibfnamefont{B.}~\bibnamefont{M\"uhlschlegel}},
  \bibinfo{journal}{Z.\ Phys.\ A} \textbf{\bibinfo{volume}{155}},
  \bibinfo{pages}{313} (\bibinfo{year}{1959}).

\bibitem[{\citenamefont{Padamsee et~al.}(1973)\citenamefont{Padamsee, Neighbor,
  and Shiffman}}]{padamsee73a}
\bibinfo{author}{\bibfnamefont{H.}~\bibnamefont{Padamsee}},
  \bibinfo{author}{\bibfnamefont{J.}~\bibnamefont{Neighbor}}, \bibnamefont{and}
  \bibinfo{author}{\bibfnamefont{C.}~\bibnamefont{Shiffman}},
  \bibinfo{journal}{J.\ Low Temp.\ Phys.} \textbf{\bibinfo{volume}{12}},
  \bibinfo{pages}{387} (\bibinfo{year}{1973}).

\bibitem[{\citenamefont{Perdew et~al.}(1996)\citenamefont{Perdew, Burke, and
  Ernzerhof}}]{perdew96a}
\bibinfo{author}{\bibfnamefont{J.~P.} \bibnamefont{Perdew}},
  \bibinfo{author}{\bibfnamefont{K.}~\bibnamefont{Burke}}, \bibnamefont{and}
  \bibinfo{author}{\bibfnamefont{M.}~\bibnamefont{Ernzerhof}},
  \bibinfo{journal}{Phys.\ Rev.\ Lett.} \textbf{\bibinfo{volume}{77}},
  \bibinfo{pages}{3865} (\bibinfo{year}{1996}).

\bibitem[{\citenamefont{Vanderbilt}(1990)}]{vanderbilt90a}
\bibinfo{author}{\bibfnamefont{D.}~\bibnamefont{Vanderbilt}},
  \bibinfo{journal}{Phys.\ Rev.\ B} \textbf{\bibinfo{volume}{41}},
  \bibinfo{pages}{7892(R)} (\bibinfo{year}{1990}).

\bibitem[{\citenamefont{Mostofi et~al.}(2014)\citenamefont{Mostofi, Yates,
  Pizzi, Lee, Souza, Vanderbilt, and Marzari}}]{mostofi14a}
\bibinfo{author}{\bibfnamefont{A.~A.} \bibnamefont{Mostofi}},
  \bibinfo{author}{\bibfnamefont{J.~R.} \bibnamefont{Yates}},
  \bibinfo{author}{\bibfnamefont{G.}~\bibnamefont{Pizzi}},
  \bibinfo{author}{\bibfnamefont{Y.-S.} \bibnamefont{Lee}},
  \bibinfo{author}{\bibfnamefont{I.}~\bibnamefont{Souza}},
  \bibinfo{author}{\bibfnamefont{D.}~\bibnamefont{Vanderbilt}},
  \bibnamefont{and} \bibinfo{author}{\bibfnamefont{N.}~\bibnamefont{Marzari}},
  \bibinfo{journal}{Comput.\ Phys.\ Commun.} \textbf{\bibinfo{volume}{185}},
  \bibinfo{pages}{2309} (\bibinfo{year}{2014}).

\bibitem[{\citenamefont{Koretsune and Arita}(2017)}]{koretsune17a}
\bibinfo{author}{\bibfnamefont{T.}~\bibnamefont{Koretsune}} \bibnamefont{and}
  \bibinfo{author}{\bibfnamefont{R.}~\bibnamefont{Arita}},
  \bibinfo{journal}{Comput.\ Phys.\ Commun.} \textbf{\bibinfo{volume}{220}},
  \bibinfo{pages}{239} (\bibinfo{year}{2017}).

\end{thebibliography}


\section*{Acknowledgments}
\noindent
The authors thank M.~S.~Bahramy for fruitful discussions and comments.
This work was partly supported by Grants-In-Aid for Scientific Research (S) from the Japan Society for the Promotion of Science (JSPS, No.\ 24224009), JST (No.\ JP16H00924), and PRESTO (JPMJPR15N5).  M.~Kriener is supported by a Grants-in-Aid for Scientific Research (C) (JSPS, No.\ 15K05140).

\end{document}


\title{Supplemental Material for \\ ``Tailoring band-structure and band-filling in a simple cubic \\(IV, III)\,--\,VI superconductor''}

\author{M.~Kriener$^{\dagger}$}
\email[corresponding author: ]{markus.kriener@riken.jp}
\author{M.~Kamitani}
\thanks{These three authors contributed equally.}
\affiliation{RIKEN Center for Emergent Matter Science (CEMS), Wako 351-0198, Japan}
\author{T.~Koretsune}
\thanks{These three authors contributed equally.}
\affiliation{RIKEN Center for Emergent Matter Science (CEMS), Wako 351-0198, Japan}
\affiliation{Department of Physics, Tohoku University, Miyagi 980-8578, Japan}
\author{R.~Arita}
\author{Y.~Taguchi}
\affiliation{RIKEN Center for Emergent Matter Science (CEMS), Wako 351-0198, Japan}
\author{Y.~Tokura}
\affiliation{RIKEN Center for Emergent Matter Science (CEMS), Wako 351-0198, Japan}
\affiliation{Department of Applied Physics and Quantum-Phase Electronics Center (QPEC), University of Tokyo, Tokyo 113-8656, Japan}

\date{\today}

\maketitle

\subsection{Sample preparation and characterization}
\label{latconst}
For the growth of \SITS\ with $x \geq 0.5$, a high-pressure synthesis method was employed. Polycrystalline samples were prepared starting from stoichiometric mixtures of high-purity elemental Sn, In, Te, and Se. The mixed powder was pressed at 5~GPa at room temperature in a cubic anvil-type high-pressure apparatus. Then the temperature was increased to $\sim 1200 - 1300^{\circ}$C and kept there for about 1~h. Before releasing the pressure, the temperature was quenched to room temperature. For comparison, we also prepared \SIT\ for $x<0.5$ by conventional melt growth. The synthesis conditions are comparatively summarized in Table~\ref{tab1}. 
\begin{table}[b]
\centering
\caption{\textbf{Synthesis condition for Sn$_{\bm 1-x}$In$_{\bm x}$Te$_{\bm 1-y}$Se$_{\bm y}$}}
\label{tab1}
\begin{tabular}{ccc}
\toprule
Composition            & $x<0.5$                         & $x\geq 0.5$                          \\ \hline \addlinespace[0.75em]
Starting               & SnTe, InTe, SnSe, InSe          & Sn, In, Te, Se                       \\ \addlinespace[0.1em]
materials              & in stoichiometric ratio         & in stoichiometric ratio              \\ \addlinespace[0.75em] \hline \addlinespace[0.75em]
Preparation            & melt growth in evacuated        & high-pressure                        \\ \addlinespace[0.1em]
method                 & sealed quartz glass tubes       & synthesis                            \\ \addlinespace[0.75em] \hline \addlinespace[0.75em]
Growth conditions      & $850 \sim 900^{\circ}$C, 12 h   & 5~GPa, $1200 \sim 1300^{\circ}$C, 1h \\ \addlinespace[0.75em] \hline \addlinespace[0.75em]
\multirow{2}*{Cooling} & furnace cooling or slow cooling & furnace-quench                       \\ \addlinespace[0.1em]
                       & (24 h) to room temperature      & to room temperature                  \\ \addlinespace[0.1em]
\bottomrule
\end{tabular}
\end{table}

\begin{figure}[h]
\centering
\includegraphics[width=8cm,clip]{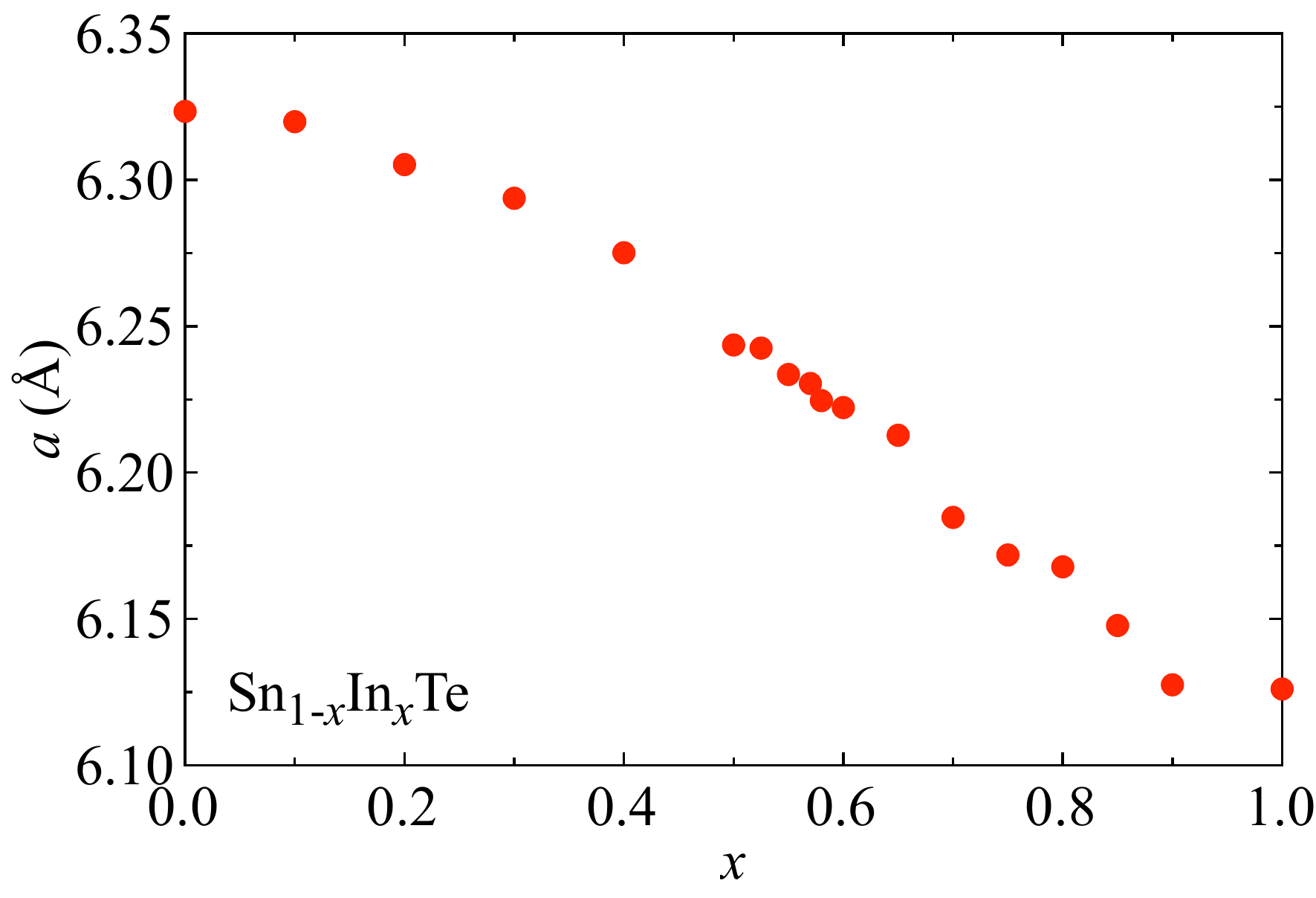}
\caption{Cubic lattice constants of \SIT.}
\label{figS1}
\end{figure}

\begin{figure}[h]
\centering
\includegraphics[width=8cm,clip]{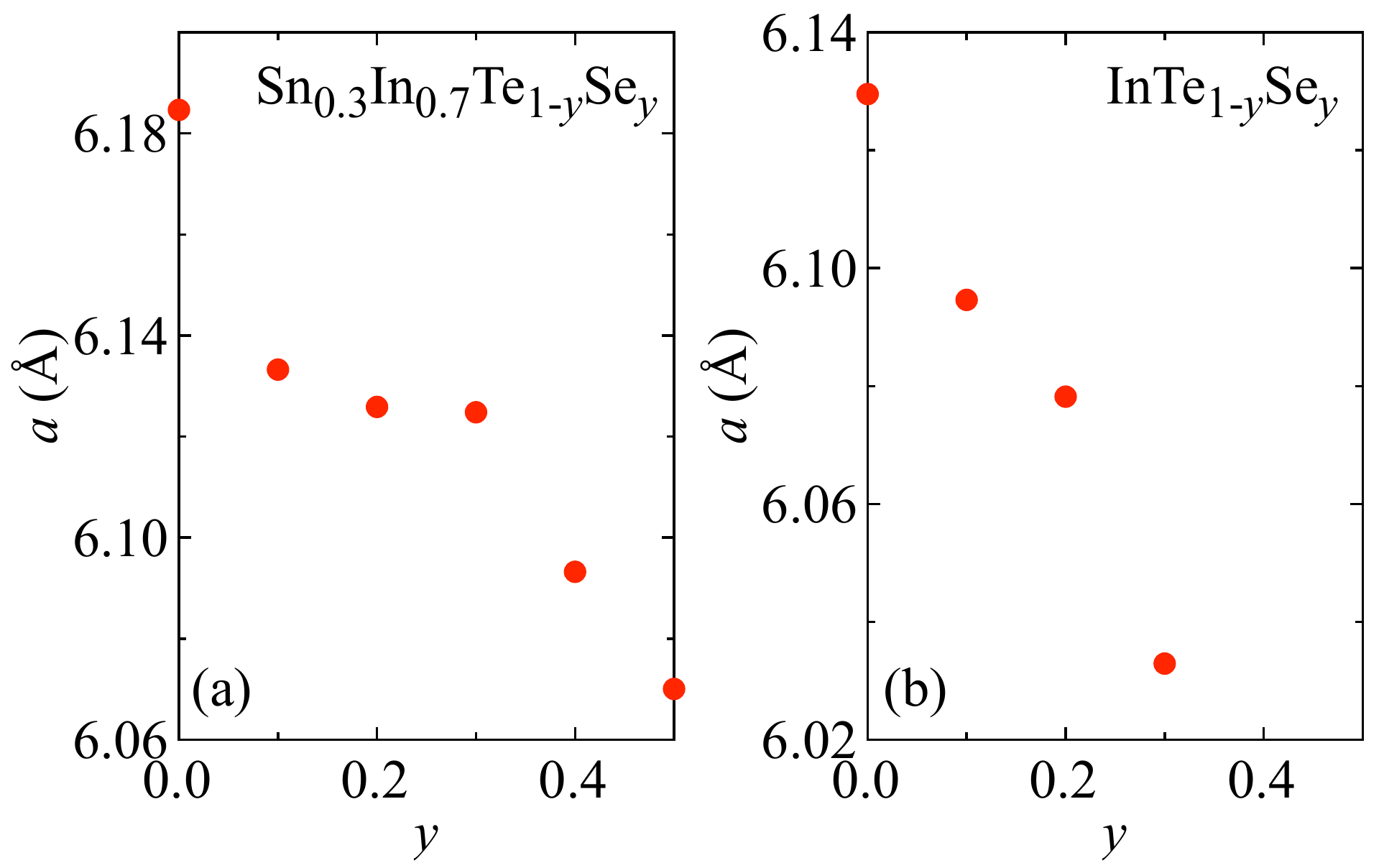}
\caption{Cubic lattice constants of \SITS\ with (a) $x = 0.7$ and (b) $x = 1$.}
\label{figS2}
\end{figure}

Powder x-ray diffraction was measured with a commercially available diffractometer (Rigaku, RINT-TTR III). Lattice constants were estimated using a commercially available software (PDXL2). As-grown material $0\leq x \leq 1$ and $0\leq y \leq 0.5$ was characterized at ambient temperature by powder x-ray diffraction with Cu-$K_{\alpha}$ radiation. Estimated cubic lattice constants $a_c$ for \SIT\ are shown in Fig.~\ref{figS1} and for \SITS\ in Fig.~\ref{figS2}(a) for $x=0.7$ and Fig.~\ref{figS2}(b) for $x=1$. In the case of \SIT, $a_c$ monotonically decreases with $x$ following roughly Vegard's law. As for \SITS, $a_c$ further decreases with $y$ for a given $x$, due to the smaller ionic radius of Se$^{2-}$ as compared to Te$^{2-}$. These observations indicate that both dopants enter substitutionally on regular lattice sites.


\clearpage
\subsection{Magnetization}
\label{MT}
\begin{figure}[h]
\centering
\includegraphics[width=10cm,clip]{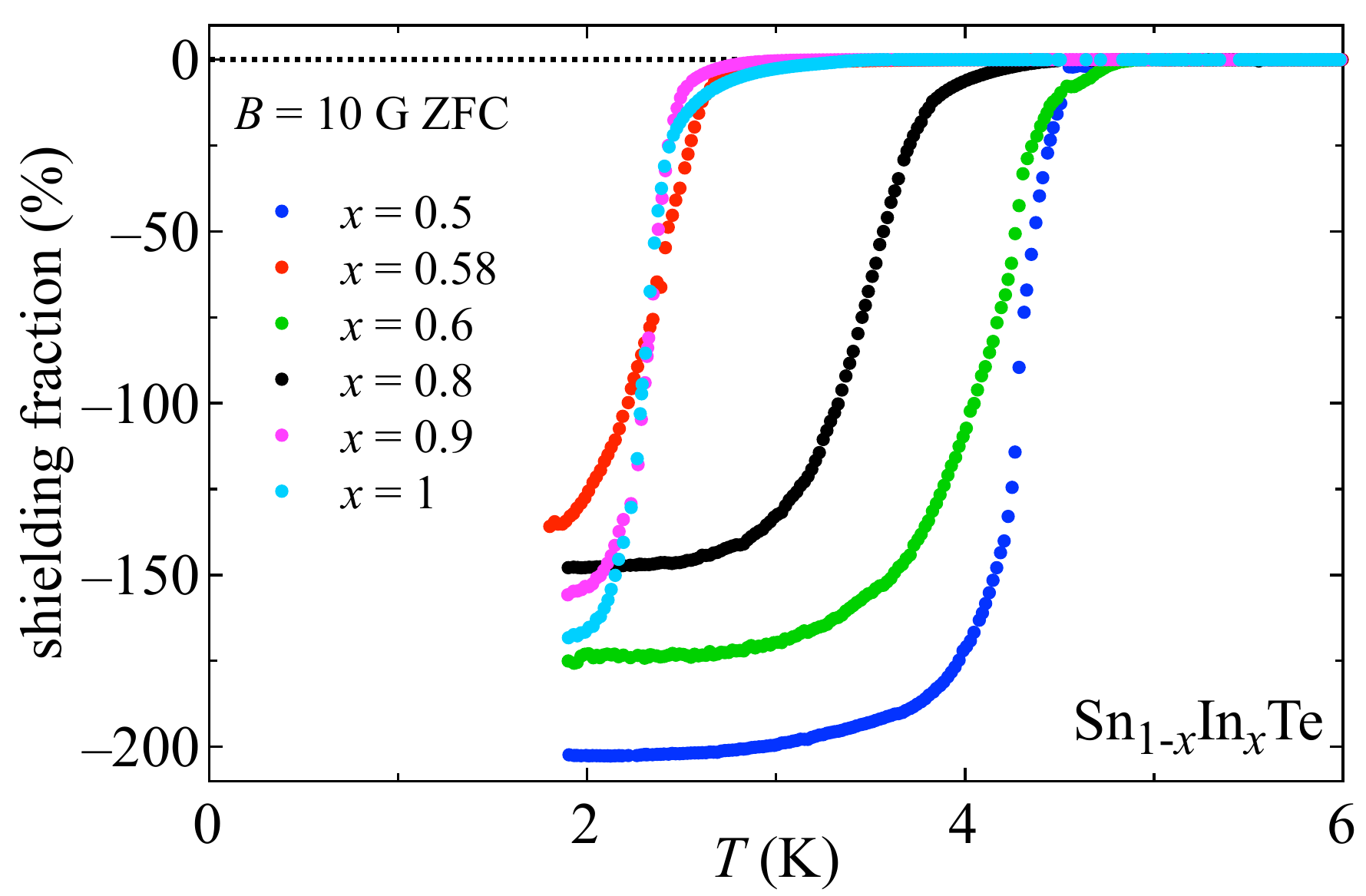}
\caption{Magnetisation of \SIT\ for $x=0.5$, 0.58, 0.6, 0.8, 0.9, and 1.0.}
\label{figS3}
\end{figure}
Magnetization was measured using a commercially available superconducting quantum interference device (MPMS, Quantum Design). Data on \SIT\ for $x=0.5$, 0.58, 0.6, 0.8, 0.9, and 1.0 are summarized in Fig.~\ref{figS3}. The data was measured upon warming after cooling in zero field (ZFC) to 1.8~K and setting $B =10$~G. Since the samples used had an irregular shape, it was not possible to correct for the demagnetization effect. As discussed in the main text, our specific-heat results indicate that all materials are bulk superconductors with volume fractions of 100\% or close to 100\% except around $x=0.58$.

\clearpage
\subsection{$\bm{T_{\rm c}}$ dependence on $\bm y$ in Sn$_{\bm{1-x}}$In$_{\bm{x}}$Te$_{\bm{1-y}}$Se$_{\bm y}$}
\label{Tcxy}
\begin{figure}[h]
\centering
\includegraphics[width=16cm,clip]{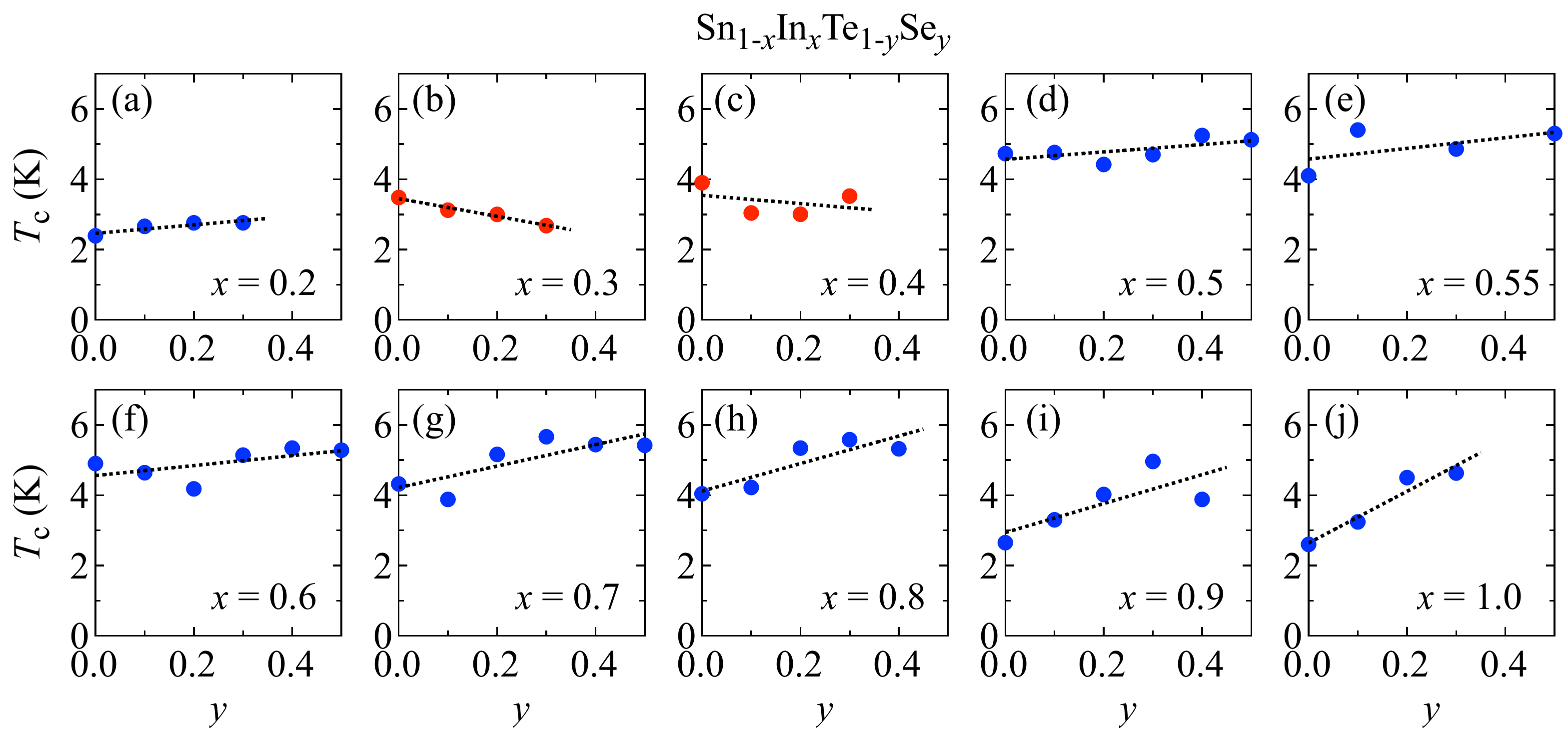}
\caption{(a)\,--\,(j), Dependence of the superconducting transition temperature \Tc\ on $y$ for fixed $x$.}
\label{figS4}
\end{figure}
Figure~\ref{figS4} summarizes the Se-doping $y$ dependence of the superconducting transition temperature \Tc\ for a fixed In concentration $0.2 \leq x \leq 1$, respectively. Blue (red) data points indicate increasing (decreasing) \Tc\ values with $y$. The dotted lines in each panel are linear fits to the data. The panels for $x=0.7$ and $x=1$ are also shown as insets in Fig.~3(c) and (d) of the main text. Apparently $d\Tc / dy$ can go up and down. 

\clearpage
\subsection{Pressure effect on $\bm{T_{\rm c}}$ in InTe}
\label{press}
\begin{figure}[h]
\centering
\includegraphics[width=10cm,clip]{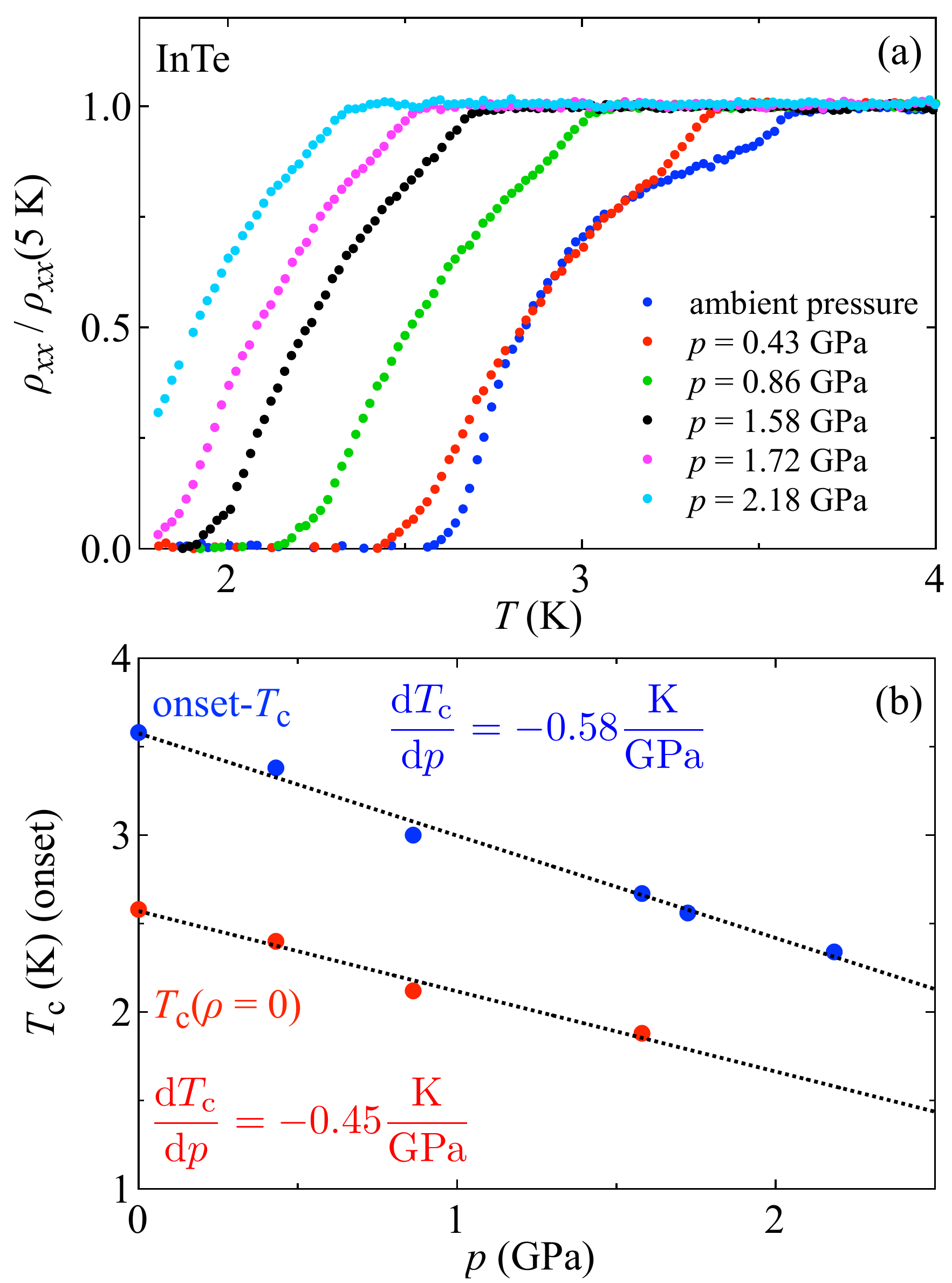}
\caption{(a) Hydrostatic pressure dependence of $\rho_{xx}$ in InTe. (b) Superconducting transition temperature \Tc\ (onset and zero resistivity) vs pressure $p$.}
\label{figS5}
\end{figure}
For resistivity measurements under high pressure, we used a standard piston cylinder clamp cell. In parallel, a Pb standard sample was also measured and the superconducting transition temperature of Pb was used to calculate the exact applied pressure. The hydrostatic pressure $p$ dependence of the resistivity $\rho_{xx}$ in InTe is shown for $p<3$~GPa in Fig.~\ref{figS5}(a). With increasing pressure \Tc\ decreases. This change is plotted against $p$ in Fig.~\ref{figS5}(b). The upper curve refers to the onset temperature of the superconducting transition and the lower curve to the temperature where zero resistivity is achieved. The latter lies below 1.8~K (lowest accessible temperature) for pressures above $\sim 1.7$~GPa. Apparently, \Tc\ decreases linearly with $p$. This behavior is often seen in conventional superconductors as, e.g., Sn, suggesting that cubic InTe is a conventional BCS-type superconductor. 

\clearpage

\subsection{Specific heat and density of states}
Specific-heat \cp\ data were also measured in a commercially available system (PPMS, Quantum Design) by employing a relaxation method. The background signal (Addenda) was measured in 0~T and 2~T prior to each sample measurement. Subsequently, the sample was mounted and measured in the same magnetic fields. The Addenda data were subtracted from the total signal (sample + Addenda) to obtain the pure sample signal. These data were analysed as follows. First, the Debye law 
\begin{equation}
\cp = \cel + \cph = \gn T + A_3 T^3 + A_5 T^5
\end{equation}
consisting of the electronic \cel\ and the phononic contribution \cph\ was fitted to the normal-state specific heat as measured in $B=2$~T. The result for \cph\ was subtracted from the total specific heat \cp. The remaining electronic specific heat $\cel = \gn T$ with the electronic specific-heat (Sommerfeld) coefficient \gn\ was further analysed in the standard weak-coupling Bardeen-Cooper-Schrieffer (BCS) theory of superconductivity. Since not all samples exhibit a superconducting volume fraction of 100\% but also consist of nonsuperconducting parts, \gn\ was rewritten as $\gn = \gs + \gres$, where \gs\ represents the superconducting and \gres\ the remaining residual normal-conducting density of states below \Tc. The latter is responsible for an additional $T$-linear contribution to \cel\ even below \Tc:
\begin{equation}
\cel = \gres T + \cel^{\rm BCS}(T).
\end{equation}
The BCS specific heat was modelled by fitting tabulated BCS data to a polynomial. This polynomial was manually adjusted to the experimental \cel\ by tuning \Tc\ and \gs\ until a satisfying description was achieved, see Refs.~[35] and [36] for details. 

The Debye temperature $\Theta_{\rm D}$ was estimated from $A_3$ via 
\begin{equation}
A_3 = \frac{12}{5}\,\frac{\pi^4 N N_{\rm A} k_{\rm B}}{\Theta_{\rm D}^3}
\end{equation}
with the number of atoms per formula unit $N=2$, the Avogadro number $N_{\rm A}$, and the Boltzmann constant $k_{\rm B}$.

According to the Drude-Sommerfeld theory the normal-state electronic specific-heat coefficient \gn\ is related to the density of states (DOS) as $\textrm{DOS} = 3\gn/(\pi^2k^2_{\rm B})$. To properly convert the experimental units and taking into account that in the BCS theory of superconductivity, the DOS is counted separately for each spin direction [37], the final formula to calculate the DOS from \gn\ is
\begin{equation}
\textrm{DOS} = \frac{3\gn}{2\pi^2k^2_{\rm B}}\cdot \frac{eV_0}{V_{\rm mol}}
\end{equation}
with the element charge $e$, the unit-cell volume $V_0$, and the molar volume $V_{\rm mol}$. For $V_0$ it is assumed that Vegard's law holds in this system which is roughly justified, cf.\ Fig.~S1.

Experimental electron-phonon constants $\lambda$ were calculated via the McMillan formula [38]
\begin{equation}\label{McMil}
\Tc=\frac{\Theta_{\rm D}}{1.45}\exp \left( -\frac{1.04(1+\lambda)}{\lambda - \mu^*(1 + 0.62 \lambda)} \right).
\end{equation}
Here we follow the usual convention and choose $\mu^* = 0.11$ for the repulsive Coulomb potential; $\Theta_{\rm D}$ is the Debye temperature.

\subsection*{Specific heat for $\bm{x = 0.2}$}
\label{Tcxy}
\begin{figure}[h]
\centering
\includegraphics[width=10cm,clip]{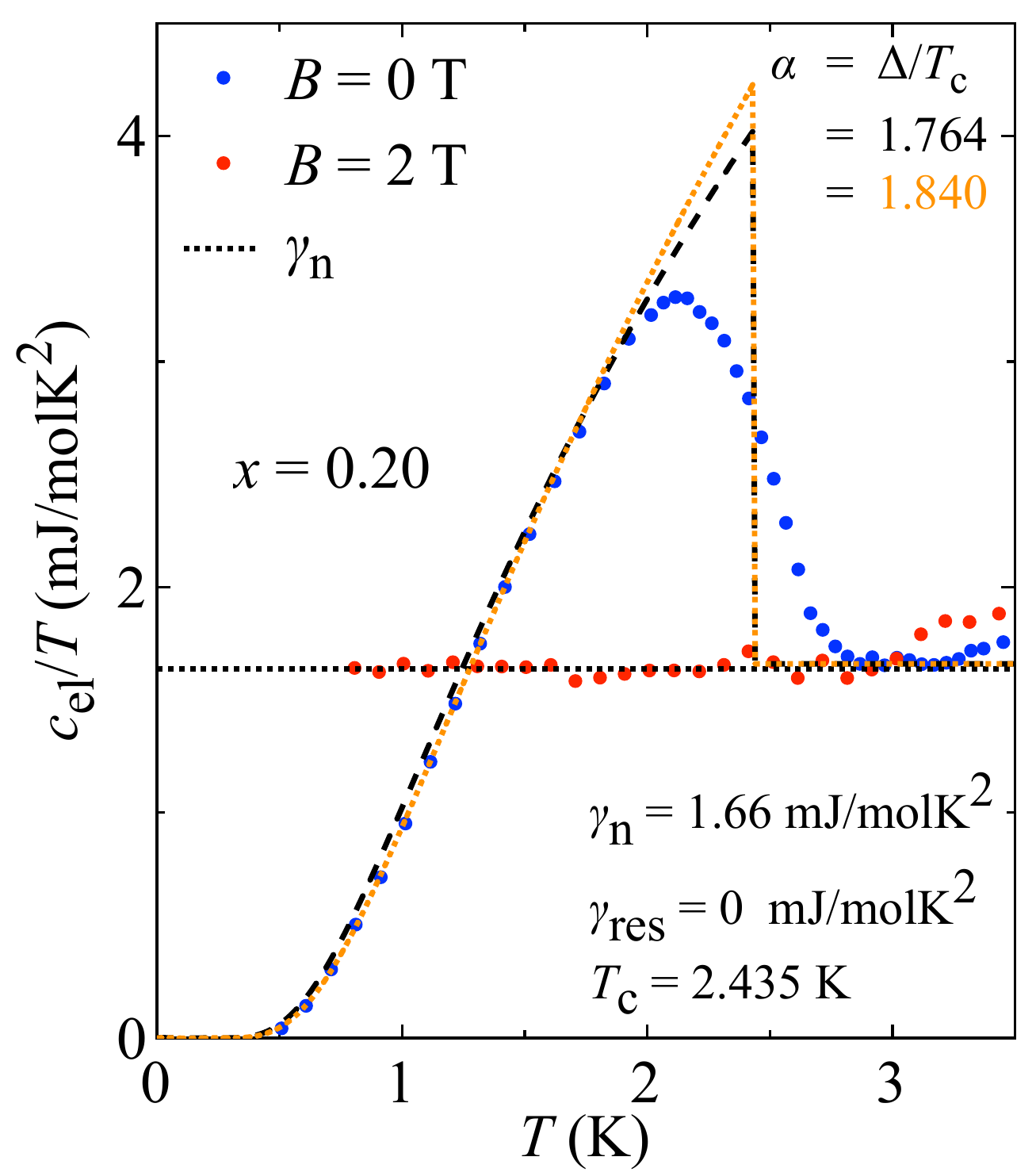}
\caption{Specific heat for $x=0.20$ and $y=0$ with weak-coupling (black dashed) and more strong-coupling (orange dotted) BCS data model calculations.}
\label{figS6}
\end{figure}
Figure~\ref{figS6} shows exemplarily the specific heat of a low-doped sample with two different BCS-based model calculations. The experimental electronic specific heat is slightly better reproduced when increasing the coupling strength $\alpha = \Delta / \Tc$ with the superconducting energy gap $\Delta$. In the standard weak-coupling BCS theory of superconductivity $\alpha = 1.764$. In the model calculation shown in orange, this parameter was increased to 1.84 to give a better description of the experimental data [39,40]. This is in agreement with earlier specific-heat reports (Refs. [12] and [18]).

\subsection{Theory} 
The DFT electronic structure calculations were performed within the generalized gradient approximation [41] as implemented in the quantum-ESPRESSO package [24]. Ultrasoft pseudopotentials [42] with cutoff energies of 50~Ry for wave functions and 1000~Ry for charge densities were used. Spin-orbit interaction is not included to our DFT calculations. The electron-phonon coupling constants were obtained using $12 \times 12 \times 12$ $k$ points for electronic structure calculations and $6 \times 6 \times 6$ $q$ points for phonon calculations. To calculate the \Tc\ for \SITS\, the electronic structures, phonon frequencies, and electron-phonon couplings for \SIT\ and \SIS\ were computed within the rigid-band approximation with lattice constants of $a = 5.80$~\AA\ for InSe and $a = 6.12$~\AA\ for InTe. Then, the phonon frequencies and electron-phonon couplings of \SIT\ and \SIS\ were linearly interpolated. The DOS of \SITS\ were obtained by interpolating the two tight-binding Hamiltonians extracted from the electronic structures of InTe and InSe [43].

For calculating \Tc, we used the McMillan-Allen-Dynes formula [25],
\begin{equation}
\Tc = \frac{\omega_{\rm log}}{1.2} \exp \left( - \frac{1.04(1+\lambda)}{\lambda - \mu^*(1 + 0.62 \lambda)} \right).
\end{equation}
As above, we chose $\mu^*=0.11$ for the pseudo Coulomb potential and $\lambda$ and $\omega_{\rm log}$ are defined using the Eliashberg function $\alpha^2F(\omega)$ as
\begin{align}
\lambda &= 2 \int d \omega \frac{\alpha^2 F(\omega)}{\omega},\label{eq:lambda}\\
\ln \omega_{\rm log} &= \frac{2}{\lambda} \int d \omega \frac{\alpha^2 F(\omega)}{\omega} \ln(\omega). \label{eq:omegalog}
\end{align}
For practical calculations of $\lambda$ and $\omega_{\rm log}$, we performed a weighted-average method for the summations over the $k$ and $q$ grid [44]. For the calculation of $\lambda$ from experimental data the conventional McMillan formula~(\ref{McMil}) was employed because the Debye temperature is experimentally accessible.

Since the DFT calculations yield the bare DOS (the ``band-structure'' DOS) while the DOS measured by specific heat is enhanced due to the electron-phonon coupling, the theoretical results were multiplied by $1+\lambda$ to correct for this difference.